# From seasons to decades: Solar radiation, cloud cover, and $CO_2$ shape young leaf phenology in a tropical forest over 26 years

Laura Lüthy[1,2,3], Colin A. Chapman[4,5,6], Patrick Lauer[1,2,7], Patrick Omeja[8], and Urs Kalbitzer[1,2,9]

[1]*Department of Biology, University of Konstanz, 78464 Konstanz, Germany*

[2]*Department for the Ecology of Animal Societies, Max Planck Institute of Animal Behavior, 78315 Radolfzell*

[3]*International Max Planck Research School for Quantitative Behaviour, Ecology and Evolution (IMPRS/QBEE); Max Planck Institute of Animal Behavior, 78315 Radolfzell, Germany*

[4]*Biology Department, Vancouver Island University, Nanaimo, British Columbia V9R 5S5, Canada*

[5]*Shaanxi Key Laboratory for Animal Conservation, Northwest University, Xi'an 710069, China*

[6]*School of Life Sciences, University of KwaZulu-Natal, Pietermaritzburg 3629, South Africa*

[7]*Max-Planck-Institute for Evolutionary Anthropology, 04103 Leipzig, Germany*

[8]*Makerere University Biological Field Station, Fort Portal, Uganda*

[9]*Centre for the Advanced Study of Collective Behaviour, University of Konstanz, 78464 Konstanz, Germany*

**Abstract**

1. Climate change is altering plant phenology globally with potential deleterious impacts on animal species and entire ecosystems, yet the long-term effects of climate change on tropical leaf production remain poorly understood.
2. We analyzed 26 years of young leaf phenology field data from Kibale National Park, Uganda, focusing on 12 tree species consumed by leaf-eating mammals. We examined seasonal and long-term patterns and how they are related to climatic variables using Bayesian hierarchical generalized additive mixed models (GAMMs).
3. The tree community and most species exhibited peaks in young leaf production during the two annual rain seasons, with seasonal changes primarily associated with diffuse light availability through solar radiation and cloud cover, as well as rainfall and minimum temperature. Long-term variations in leaf production was primarily linked to long-term changes in atmospheric $CO_2$, solar radiation, and cloud cover.
4. Our results support the role of $CO_2$ fertilization, though decreasing levels of solar radiation resulting from the ending of the recent solar cycle may be slowing this effect.
5. Synthesis: This study highlights the critical role of diffuse light, solar radiation, and the solar cycle in predicting tropical leaf production, emphasizing that interpretations of greening trends must consider solar radiation alongside atmospheric $CO_2$ levels. Furthermore, our findings emphasize the complex relationship between climate and young leaf phenology, highlighting the importance of integrating species-specific long-term data to better understand the effects of climate change on food availability for tropical folivores and tropical forest ecosystems in general.

## Introduction

Climate change impacts ecosystems worldwide, including tropical ecosystems, which are home to the majority of Earth's biodiversity and are especially vulnerable (Barlow et al., 2018). Temporal patterns in plant productivity, such as variation in the production of leaves, flowers, or fruit (i.e. plant phenology), is a key process linking climate to ecosystem dynamics (Abernethy et al., 2018). For instance, seasonal leaf flushing drives canopy productivity and food availability for higher trophic levels (Lauer et al., 2025; Matsuda et al., 2020; van Schaik et al., 1993). The Intergovernmental Panel on Climate Change (IPCC) has emphasized that phenology is one of the simplest processes to track changes in the ecology of species in response to climate change (Rosenzweig et al., 2007). While much research on climate change impacts on phenology has focused on temperate forests and on flowering and fruiting in the tropics, comparatively little attention has been given to how climate change influences leaf phenology in tropical forests (Davis et al., 2022; Piao et al., 2019). This is despite the ecological importance of leaves and particularly young leaves as a key food resource for many animal taxa in these ecosystems (Chapman & Chapman, 2002; Coley & Barone, 1996; van Schaik et al., 1993). However, obtaining long-term phenological datasets, especially with species-specific information relevant for understanding ecosystem processes, is difficult. Nevertheless, these long-term patterns are required to identify environmental drivers of changing phenology patterns (Sullivan et al., 2024).

Plants often exhibit seasonal (intra-annual) and long-term (inter-annual to decadal) patterns of phenology, which may respond differently to climatic variability and be governed by different mechanisms. For example, across the tropics, seasonal fluctuations in young leaf availability has been associated with seasonal variability in rainfall and solar radiation (van Schaik et al., 1993; Wright & van Schaik, 1994), and peaks in flushing often coincide with annual periods of high solar radiation when water is not a limiting factor (van Schaik et al., 1993; Wagner et al., 2017; Wright & van Schaik, 1994; Xiao et al., 2006; Yang et al., 2021). Another relevant factor is cloud cover, which reduces total solar radiation but simultaneously increases the amount of diffuse solar radiation reaching plant canopies and the lower leaves in a forest. Thus, increased cloud cover in combination with high solar radiation can lead to increased plant productivity and leaf flushing (Barone, 1998; Durand et al., 2021; Gu et al., 2002; Wild et al., 2012; Williams et al., 2016). However, the importance of rainfall, solar radiation, and diffuse solar radiation on leaf production over longer timescales remains largely unexplored. Given that long-term variation in solar radiation is associated with the 11-year solar activity cycles (Hathaway, 2015; Hempelmann & Weber, 2012), these cycles may also lead to long-term variation in forest productivity, which is potentially an important gap in our understanding of tropical ecology.

Atmospheric $CO_2$ is another critical resource for plant productivity, and experiments have shown that $CO_2$ enrichment can enhance photosynthesis and water-use efficiency in trees (Ainsworth & Rogers, 2007). While atmospheric $CO_2$ fluctuates seasonally, this is largely driven by terrestrial biosphere activity (Gulev et al., 2021), making it difficult to determine whether short-term fluctuations drive young leaf production or instead reflect vegetation productivity. In contrast, the sustained long-term increase in atmospheric $CO_2$ has been proposed as a major contributor to the global greening trend observed over recent decades through $CO_2$ fertilization (Piao et al., 2020; Zhu et al., 2016), although evidence suggests a weakening of the $CO_2$ fertilization effect from 1982 to 2015 (Wang et al., 2020).

Temperature also influences leaf production, as moderate increases in temperature can accelerate photosynthesis, but extremely high temperatures can raise vapour pressure deficits and impose physiological stress, potentially leading to declines in photosynthetic rates (Lewis et al., 2004). In temperate forests, seasonal increases in minimum temperature often act as a major cue for leaf flushing, and in tropical forests, minimum temperature and humidity have been linked to leaf flushing (Jose et al., 2025; Kobayashi et al., 2020). Temperature may influence leaf phenology differently across seasonal and long-term timescales, given the contrasting mechanisms of minimum and maximum temperature impacting plant productivity.

Finally, large-scale long-term climate phenomena, such as the El Niño Southern Oscillation (ENSO), influence rainfall, temperature, cloud cover, and other climatic factors, and may thereby affect tree phenology as well (Jin et al., 2025). For example, ENSO has been identified as an important driver of tropical fruit phenology and vegetation growth (Chapman et al., 2018; Jin et al., 2025). In eastern Africa, the Indian Ocean Dipole (IOD) also plays a major role in shaping inter-annual climate variability (Marchant et al., 2007) and vegetation growth (Jin et al., 2025), and may therefore contribute to long-term phenological patterns.

To obtain a broader understanding of the impact of climate factors on tropical trees and ecosystems, we investigate seasonal and long-term young leaf phenology patterns, and how these patterns are associated with climatic factors using 26 years of field-based phenological data collected in Kibale National Park, Uganda (hereafter Kibale). Kibale hosts a moist tropical evergreen forest which serves as a conservation refuge for many species (Chapman & Lambert, 2000). At the study location, the forest experiences annual rainfall of approximate 1,646 mm with two annual rain seasons, and monthly minimum and maximum temperatures ranging from 15.06°C to 28.9°C, which have increased by approximately 1°C from 1970 to 2020 (Chapman et al., 2021). We focus on 12 tree species whose young leaf phenology patterns are highly relevant for the ecosystem in Kibale as they present the main food resources for an important folivore in Kibale, the endangered Ugandan red colobus monkey (*Piliocolobus tephrosceles)* (Lauer et al., 2025).

We have two specific aims. First, we quantify the community-wide (i.e., all 12 tree species) and species-specific seasonal and long-term patterns of young leaf production (Aim 1). Second, we evaluate which climatic drivers best explain the documented patterns (Aim 2). For Aim 1, we hypothesize that young leaf production in general is seasonal and varies across years given the observed seasonal and long-term climatic variability in Kibale. For Aim 2, we formulate a set of hypotheses based on established physiological mechanisms and the observed links between climate and leaf production outlined above. We distinguish seasonal from long-term impact of climatic drivers and then evaluate and compare our hypotheses using a model comparison approach. At seasonal timescales, we hypothesize that young leaf production is primarily driven by short-term intra-annual fluctuations in resource availability for photosynthesis, particularly water and light, with minimum temperature potentially acting as an additional physiological cue. We evaluate five hypotheses (**Table 1**) that do not include strongly correlated variables (e.g., rainfall and cloud cover; maximum and minimum temperature) and our model comparison approach allows us to assess whether models that include, for example, either rainfall or cloud cover are more informative predicting young leaf production. At long-term timescales, we hypothesize that gradual changes in atmospheric conditions and large-scale climate variability shape patterns of young leaf production and evaluated seven different hypotheses (**Table 1**). Here, we did not combine ENSO and IOD with other climatic variables because we assumed that if these phenomena are linked to phenology, such an association would occur through other climatic variables.

***Table 1 Hypotheses for the effects of climate variables on seasonal and long-term young leaf production.*** *Seasonal hypotheses ($H1–H5_{seasonal}$) propose mechanisms explaining short-term intra-annual variation in leaf flushing. Long-term hypotheses ($H1–H7_{long\text{-}term}$) propose explanations for long-term trends in young leaf production. Signs in brackets (+,-) indicate whether we hypothesize a positive of negative relationship.*

| Hypotheses to explain seasonal patterns | | |
|---|---|---|
| $H1_{seasonal}$ | Solar radiation (+) and rainfall (+) | Seasonal leaf production is primarily limited by water and energy availability. |
| $H2_{seasonal}$ | Solar radiation (+), rainfall (+) and minimum temperature (+) | In addition to water and energy limitations, an increase in minimum temperature acts as a cue for seasonal young leaf production. |
| $H3_{seasonal}$ | Rainfall (+) and minimum temperature (+) | Only water is a limiting resource, and increases in minimum temperature act as a seasonal cue. |
| $H4_{seasonal}$ | Cloud cover (+) and solar radiation (+) | High solar radiation and cloud cover lead to an increase in diffuse light, which increases leaf production. |
| $H5_{seasonal}$ | Cloud cover (+), solar radiation (+) and minimum temperature (+) | Diffuse light availability drives leaf production, and minimum temperature acts as additional temporal cue. |
| Hypotheses to explain long-term patterns | | |
| $H1_{long\text{-}term}$ | Solar radiation (+) and rainfall (+) | Long-term leaf production is primarily limited by joint water and energy availability. |
| $H2_{long\text{-}term}$ | Cloud cover (+) and solar radiation (+) | Long-term increases in solar radiation and diffuse light lead to higher long-term leaf production. |
| $H3_{long\text{-}term}$ | Solar radiation (+), cloud cover (+) and $CO_2$ (+) | $CO_2$ in addition to solar radiation and diffuse light condition increases young leaf production reflecting the $CO_2$ fertilization effect. |
| $H4_{long\text{-}term}$ | Solar radiation (+), $CO_2$ (+) and rainfall (+) | Joint changes in solar radiation, rainfall, and atmospheric $CO_2$ influence leaf production through their combined effects on photosynthesis and water-use efficiency. |
| $H5_{long\text{-}term}$ | Rainfall (+) and maximum temperature (-) | Long-term rainfall increases lead to higher, whereas maximum temperature increases lead to lower young leaf production. |
| $H6_{long\text{-}term}$ | ENSO (+) | ENSO, a large-scale climate oscillation affecting multiple climate variables best predicts the long-term variation in young leaf production |
| $H7_{long\text{-}term}$ | IOD (+) | IOD, a large-scale climate oscillation affecting climate in East Africa best predicts long-term variation in young leaf production, better than ENSO. |

## Materials and Methods

### Study site

Data were collected in Kibale National Park, a mid-altitude, moist-evergreen forest in western Uganda (0°13'–0°41'N and 30°19'–30°32'E) (Chapman et al., 2010, 2021). The field site is situated near the Makerere University Biological Field Station in Forestry Compartment K30, a section of the forest that has never been commercially logged except for a few large stems logged prior to 1970 (Chapman et al., 2010). This area has previously been described as a relatively undisturbed old forest (Struhsaker, 1997). Over the last 50+ years the forest did not receive any major disruption or changes in community composition (Chapman et al., 2021; Chapman unpublished 2025 data).

Phenology data

Phenology data for young leaves were collected approximately every 4 weeks between February 1999 and December 2024, only paused from April to June 2020 due to the Covid-19 pandemic. Each record was assigned to a specific month. For our analysis, we included data from 12 tree species that represent the most consumed species by folivorous red colobus in the area, making up ~50% of their diet (*Albizia grandibracteata, Celtis africana, Celtis gomphophylla, Dombeya kirkii, Funtumia africana, Macaranga schweinfurthii, Millettia dura, Parinari excelsa, Prunus africana, Strombosia scheffleri, Trilepisium madagascariense, Vepris nobilis)* (Lauer et al., 2025). For each record, the tree crown of each tree was visually examined from the ground and given a young leaf score (YL) from 0 to 6 relative to the maximum young leaf cover, with zero representing 0% of the crown bearing young (i.e. newly emerged) leaves, and 6 representing 100% of maximum young leaf cover. The score therefore quantifies the proportion of the crown with young/newly emerged, still soft leaves that are visually distinguishable from mature leaves based on characteristics such as lighter coloration, softer texture, and incomplete expansion. This scoring system is similar to other phenology studies (Luse Belanganayi et al., 2025; Williams et al., 1997) and allows comparison of phenological patterns across individuals and species. All observations have been conducted by the same long-term field assistants since 1999, with Tusiime Laurence leading the data collection for the entire study. In total, we included 33,956 observations in our analysis, with an average number of 10 individual trees per species and month (range = 1 - 26), and an average of 119 trees every month (range = 86 - 248).

Abiotic variables

Daily rainfall data were collected at the field site using a circular rain gauge, summarized per month in mm (hereafter *Rain*). Since temperature data collected at the same weather station were impacted by relocations and changes of thermometers (described in Chapman et al., 2021), we obtained monthly maximum temperature (hereafter *Tmax*, in °C) and minimum temperature (hereafter *Tmin*, in °C) from the CRU TS v. 4.09 dataset (https://crudata.uea.ac.uk/cru/data/hrg/cru_ts_4.09/) (Harris et al., 2020). Monthly downward surface shortwave radiation data (hereafter solar radiation or *Srad*, in $W/m^2$) were derived from the TerraClimate database accessed through Google Earth Engine (Abatzoglou et al., 2018; Gorelick et al., 2017). Monthly cloud fraction data (hereafter cloud cover or *Cloud*, in %) represents the percentage of cloudy pixels in a grid cell, were derived from the satellite-based data record CLARA-A3, created by the EUMETSAT's Satellite Application Facility on Climate Monitoring (CM SAF) (Karlsson et al., 2023). Additionally, we included atmospheric $CO_2$ (in ppm) from globally averaged marine surface monthly mean data (Lan et al., 2024).

For the intensity of the El Niño Southern Oscillation (ENSO), we included the monthly timeseries of the Multivariable ENSO Index Version 2 (MEI.v2, Kobayashi et al., 2015). The Indian Ocean Dipole (IOD) is represented by the anomalous sea surface temperature gradient between the western equatorial Indian Ocean and southeastern equatorial Indian Ocean, the Dipole Mode Index (DMI; (Saji & Yamagata, 2003). Both the MEI.v2 and DMI were downloaded from the NOAA Physical Sciences Laboratory (https://psl.noaa.gov/data/timeseries/month/DS/DMI/,https://psl.noaa.gov/enso/mei/).

Statistical modelling

Phenological and climate patterns are inherently time-dependent, non-linear, and cyclical, often exhibiting multiple peaks within a year, and long-term changes across years. Additionally, phenological field data sets commonly consist of repeated observations of the same individual trees and species, leading to a hierarchical data structure. To accommodate these characteristics, we analysed our phenology and climate time series using generalized additive mixed models (GAMMs), which provide a flexible framework to model complex, non-linear relationships while accommodating circular seasonal patterns and hierarchical data structures (Polansky & Robbins, 2013; Simpson, 2018). An additional benefit of GAMMs is that they can handle missing and unevenly spaced observations and prevent overfitting by using automatic selection methods for the complexity of splines with smoothness penalty terms (Simpson, 2018).

Our modelling approach is conceptually aligned with a time-series decomposition used in other ecological studies (Polansky & Robbins, 2013; Rojo et al., 2017; Verbesselt et al., 2010), in which an observed signal is separated into different temporal components. Here, we decompose the phenological time series into two additive components using spline smoothers to investigate phenological patterns at seasonal and long-term scales (Aim 1) and how climatic variables are linked to these phenological seasonal and long-term patterns (Aim 2).

We implemented all models in a Bayesian framework in Stan (Stan Development Team, 2023), which provides flexibility to analyze and visualize the detected patterns considering the uncertainty of model estimates, and offers powerful tools for model diagnostics (McElreath, 2018). We use estimated posterior mean values and 95% credible intervals (CrI) derived from posterior distribution to determine whether differences are meaningful. If mean values and their 95% CrI of estimated coefficients are not intersecting with 0, we label them as 'clearly' positively/negatively associated with the outcome variable. Similarly, we compare peak and low values by considering their posterior mean values and 95% CrIs.

*Phenological patterns (Aim 1)*

For the community (all species together) and each species separately, we modelled young leaf scores (*YL*) as a binomial outcome variable with six trials (because scores range from 0 to 6):

$$YL_i \sim Binomial(trials\ =\ 6, p_i) \text{ (Eq. 1)}$$

$$logit(p_i) = \alpha + SSp_{Month} + TSp_{MonthSinceStart} + random\ effects$$

Here, $p$ represents the estimated probability of leaf production and $\alpha$ the intercept. The term $SSp_{Month}$ is the s*easonal spline term (SSp)* where we used a cyclic cubic regression spline with month of the year as the predictor to capture within-year phenological seasonality. We specified a basis dimension of 12 (one per month), providing sufficient flexibility to describe seasonal patterns. Using a cyclic spline ensured continuity between December and January. The term $TSp_{MonthSinceStart}$ represents the *trend spline term (TSp)*, where we used a non-cyclic cubic regression spline with months since the first observation as the predictor. We specified a basis dimension of 26 (one per year), allowing the spline to represent gradual changes over time. For both splines, smoothness was controlled by penalization during model fitting to prevent overfitting. For the random effects, we always included a unique identifier for each tree (*TreeID*) as random intercept to account for the clustered nature of data (i.e. repeated observations from the same tree) and potentially varying baseline probabilities for different trees. For the community model, we also included the species of each tree as an additional random intercept.

To quantitatively investigate phenological patterns, we then derived several metrics from the seasonal and long-term prediction curves of the models. These curves are based on the intercept and either the *SSp* or *TSp* term, and predict mean young leaf scores ($YL_{mean}$) for different values of *SSp* or *TSp*. We quantified seasonal and long-term peaks and lows by identifying local maxima and minima of $YL_{mean}$, and the amplitude as the difference $\Delta YL_{mean}$ between the global peak and global low. For the seasonal curves, we first determined the number and locations of distinct peaks and lows that did not overlap the CrI of preceding and following lows/peaks. We then quantified the strength of seasonality using the amplitude between predicted global minimum and maximum $YL_{mean}$, classifying amplitudes of 0–1 as very weak, 1–2 as weak, >2 as intermediate, and values above 3 as strong seasonality (maximum possible amplitude = 6 $\Delta YL_{mean}$). Long-term phenological patterns were classified using a similar approach. We first quantified meaningful peaks and lows where the predicted peak/low did not overlap with the 95% CrI of the preceding low/peak. We then defined overall long-term variability as the amplitude between the global predicted maximum and minimum young leaf scores.

*Testing hypotheses about the associations between climatic variables and young leaf phenology (Aim 2)*

To investigate how seasonal and long-term patterns of young leaf productions are linked to climatic variables at the community and species level, we constructed models reflecting our different hypotheses (Table 1) and compared them using a model comparison approach. We did this for the community and each species separately. More specifically, to investigate how climate variables are linked to seasonal patterns, we replaced the $SSp_{Month}$ term in the phenology model (Equation 1) with adjusted seasonal climate variables ($CVs_{seasonal}$; *details below*) specified by each hypothesis (Equation 2a). To evaluate which climate variables best explain long-term phenological patterns, we replaced the $TSp_{MonthSinceStart}$ term by the adjusted long-term trend climate variables ($CVs_{long\text{-}term}$, Equation 2b):

$$YL_i \sim Binomial(trials\ =\ 6, p_i)$$

$$logit(p_i) = \alpha + CVs_{seasonal} + TSp_{MonthSinceStart} + random\ effects \text{ (Eq 2a)}$$

$$logit(p_i) = \alpha + CVs_{long-term} + \ SSp_{Month} + random\ effects \text{ (Eq. 2b)}$$

Replacing the *SSp* allowed us to test which climate variable best explains phenological seasonality, whereas replacing the *TSp* allowed us to test which climate variable best explains phenological long-term change.

To isolate the seasonal and long-term components of each climate variable ($CVs_{seasonal}$ and $CVs_{long\text{-}term}$), we fitted GAMMs analogous to those used for the phenology data to each climate variable (*Tmax, Tmin, Rain, Cloud, Srad,* $CO_2$*, ENSO* and *IOD*), but using a normal distribution as likelihood function:

$$CV_i \sim Normal(\mu_i, \sigma)$$

$$\mu_i = \alpha + SSp_{Month} + TSp_{MonthSinceStart} \text{ (Eq. 3)}$$

Here, $CV$ represents the observed value of the climate variable, $\mu$ is its expected mean value, $\sigma$ is the standard deviation and $\alpha$ is the intercept. The seasonal spline term $SSp_{Month}$ and the trend spline term $TSp_{MonthSinceStart}$ were specified identically to those used in the phenology models. Model results for all the climate models are detailed in the Supporting Information (Fig.S1, Fig. S2).

We used these model results in a manner analogous to a time-series decomposition to derive climate values either reflecting only seasonal ($CVs_{seasonal}$) or only long-term variability ($CVs_{long\text{-}term}$). We derived the adjusted seasonal climate variables $CVs_{seasonal}$ by first predicting the long-term trend mean values (using the posterior samples for the equation $\alpha + TSp_{MonthSinceStart}$), and subtracting these estimates from the observed climate values:

$$CV_{seasonal_i} = CV_i - mean(Normal(\alpha + TSp_{MonthSinceStart,i}, \sigma)) \text{ (Eq. 4a)}$$

For the adjusted long-term trend climate variables $CVs_{long\text{-}term}$, we subtracted predictions from the seasonal mean values ($\alpha + SSp_{Month}$) from the observed climate values:

$$CV_{long-term_i} = CV_i - mean(Normal(\alpha + SSp_{Month,i}, \sigma)) \text{ (Eq. 4b)}$$

Because phenological observations most likely depend on preceding rather than current climate conditions, we used climate values from preceding months. For seasonal predictions, we used the $CV_{seasonal}$ values from the previous month. For long-term predictions, we used the rolling mean of the $CV_{long\text{-}term}$ values of the last three months for each climate variable except for rainfall, where we use the rolling sum (i.e., cumulative rainfall). Climate variables were tested for pairwise correlation using the Pearson correlation coefficient (Pearson, 1895), and climate variables that were highly correlated (e.g. *Rain* + *Cloud; Tmax* + *Tmin; Tmax* + $CO_2$) were not used in the same model (SI Fig S3, S4).

All phenology models were fitted in a Bayesian framework using weakly informative priors. Fixed effects were assigned $\beta \sim Normal(mean = 0, sd = 1.5)$, the intercept $\alpha \sim Normal(mean = -2, sd = 2)$, and standard deviations of random effects and spline smooths $\sigma \sim Exponential\,(1)$ for tree-level and seasonal spline smooths, and $\sigma \sim Exponential\,(0.1)$ for the long-term trend spline. Climate models were also fitted in a Bayesian framework but used the default priors in *brms* version 2.20.12 (Bürkner, 2017; Vehtari et al., 2017). Model convergence and sampling quality were assessed using the rank-normalized potential scale reduction statistic (R-hat < 1.05), effective sample sizes (all bulk and tail ESS > 350 for population-level parameters), and the proportion of divergent transitions (< 5% for all models), indicating reliable posterior estimation (Stan Development Team, 2023; Vehtari et al., 2021).

*Model comparison*

To identify the hypotheses and climate variables that best predict the observed seasonal and long-term phenological patterns, respectively, we compared the different models using the leave-one-out cross-validation information criterion (LOO-IT) based on the expected log pointwise predictive density (ELPD-LOO), implemented in *brms* version 2.20.12 (Bürkner, 2017; Vehtari et al., 2017). Models were considered equally informative when their ELPD-LOO values differed by less than two standard errors of this difference.

### Use of AI in manuscript preparation

All sections of this manuscript were first structured and written by the authors without the use of AI tools. Only after completing a full draft did the authors use ChatGPT (OpenAI, 2025) to help refine language and improve readability. All AI-assisted edits were carefully reviewed and revised by the authors to ensure accuracy, clarity, and appropriateness.

### Ethics

The research described here was purely observational and was approved by the Uganda Wildlife Authority (UWA) and the Uganda National Council for Science and Technology (UNCST).

## Results

### Seasonal patterns of young leaf production

Community-wide young leaf production exhibited a peak in April ($YL_{mean}$ = 2.29, 95% CrI [2.06, 2.51]) which was clearly higher (i.e., not intersecting with 95% CrIs) than the corresponding minima in August ($YL_{mean}$ = 1.69, 95% CrI [1.49, 1.88]). Nevertheless, the seasonal amplitude at the community level was modest ($\Delta YL_{mean}$ = 0.60).

Across species, the occurrence of two or one annual maxima represented a consistent feature, with five species exhibiting two distinct peaks, and four species one peak (Fig. 1, Table 2). In addition, two species displayed only a single distinct low without a corresponding peak, indicating a dip in young leaf production rather than peaks in flushing. Only one species (*Strombosia scheffleri*) showed no distinct seasonal extrema.

Peak timing varied across species, but several species exhibited peaks in April (with additional ones in March and May) and a second cluster in September/October, which broadly coincided with the rainy seasons (Mar-May and Sep-Nov; Fig S1). Lows were most frequent in January/February and July/August, which coincided with the dry seasons (Dec-Feb and Jun-Aug).

The magnitude of seasonality varied substantially among species (Table 2): one species exhibited intermediate seasonality ($\Delta YL_{mean} > 2$), two showed weak seasonality ($\Delta YL_{mean}$ 1–2), whereas the majority displayed very weak seasonality ($\Delta YL_{mean} < 1$). Importantly, even species with low seasonal amplitudes showed seasonal patterns with clear seasonal peaks and lows, and therefore distinct periods of higher and lower young leaf production.

***Table 2 Seasonal amplitude and timing of seasonal extrema in young leaf production for the community and individual species****. Amplitude represents the difference (ΔYLmean) between the global maximum and minimum predicted young leaf scores ($YL_{mean}$; original scale 0-6)) predicted by the seasonal smooth term (+ intercept) derived from Bayesian generalized additive mixed models. For each peak and low, the month of occurrence and the respective $YL_{mean}$ value are given, with 95% credible intervals in brackets. Dashes indicate that no distinct seasonal peak or low was identified for that species. Global peaks and lows are highlighted in bold.*

| Species | Amplitude | Peaks | Lows |
|---|---|---|---|
| *Community* | 0.60 | **Apr: 2.29 [2.06, 2.51]** | **Aug: 1.69 [1.49, 1.88]** |
| *Albizia grandibracteata* | 1.19 | **Apr: 2.49 [2.32, 2.67]** | **Jan: 1.3 [1.17, 1.44]** |
| *Celtis africana* | 0.70 | **Sep: 2.5 [2.28, 2.71]** | **Aug: 1.8 [1.6, 2.01]** |
| *Celtis gomphophylla* | 2.38 | **Apr: 3.26 [3.08, 3.43]**<br>Oct: 1.74 [1.59, 1.89] | **Jan: 0.88 [0.77, 1]**<br>Aug: 1.42 [1.3, 1.55] |
| *Dombeya kirkii* | 1.37 | Apr: 2.91 [2.72, 3.1]<br>**Oct: 3.07 [2.87, 3.26]** | Feb: 1.72 [1.54, 1.88]<br>**Aug: 1.7 [1.53, 1.87]** |
| *Funtumia africana* | 0.47 | **Mar: 2.03 [1.91, 2.15]** | **Jul: 1.55 [1.44, 1.67]** |
| *Macaranga schweinfurthii* | 0.52 | May: 2.23 [2.08, 2.4]<br>**Dec: 2.31 [2.15, 2.49]** | Feb: 1.85 [1.66, 2.03]<br>**Aug: 1.8 [1.62, 1.96]** |
| *Millettia dura* | 0.95 | Apr: 2.05 [1.83, 2.28]<br>**Oct: 2.09 [1.86, 2.32]** | Jan: 1.18 [1.01, 1.35]<br>**Jul: 1.15 [0.99, 1.31]** |
| *Parinari excelsa* | 0.80 | **Feb: 2.98 [2.82, 3.15]** | **Jun: 2.18 [2.02, 2.35]** |
| *Prunus africana* | 0.53 | - | **May: 1.77 [1.59, 1.95]** |
| *Strombosia scheffleri* | 0.13 | - | - |
| *Trilepisium madagascariense* | 0.36 | - | **Jul: 1.88 [1.75, 2.02]** |
| *Vepris nobilis* | 0.46 | May: 1.93 [1.78, 2.08]<br>**Nov: 2 [1.87, 2.14]** | Feb: 1.58 [1.45, 1.71]<br>**Jul: 1.53 [1.42, 1.66]** |

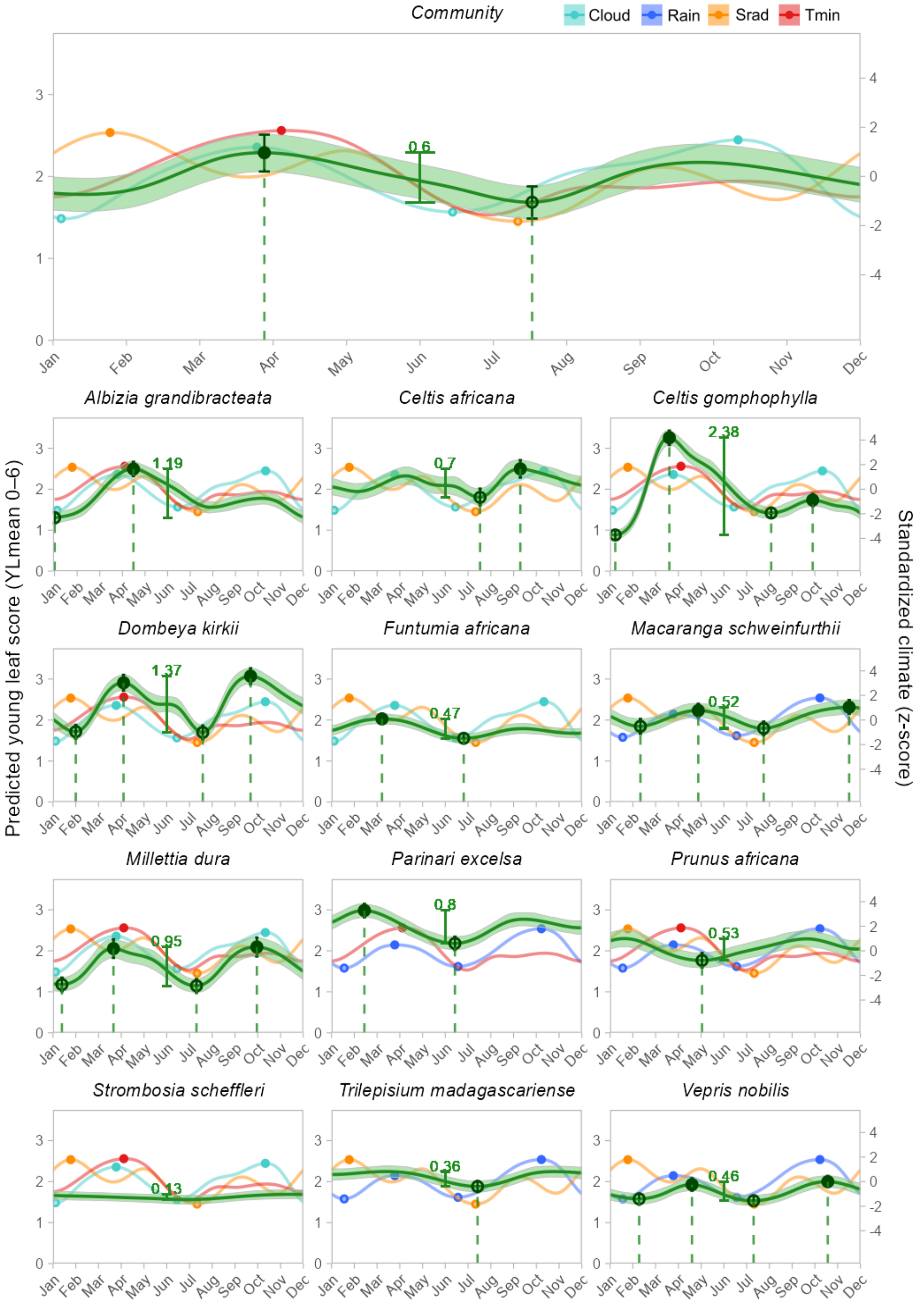

***Figure 1 Seasonal patterns of young leaf production across the community and individual species.*** *Shown in green are predictions for young leaf scores (YLmean; original scale 0 - 6) based on the seasonal smooth term (SSp + intercept) from Bayesian generalized additive mixed models. Solid lines represent posterior means, and shaded bands indicate 95% credible intervals (CrI). Points mark the timing of clear seasonal extrema (peaks and lows), with dashed vertical lines indicating their positions within the annual cycle. Green, vertical segments illustrate the seasonal amplitude (ΔYLmean, difference between global peaks and lows). The seasonal smooth term (SSp) of the climate variables included in the most informative model for each species is presented in the respective species plot. For visual comparison across variables, climate variables were standardized (z-scaled) by subtracting the mean and dividing by the standard deviation. Only extrema with non-overlapping 95% credible intervals were marked with points.*

Association between seasonal patterns of climatic variables and young leaf production

Across community- and species-level analyses for seasonal climate effects, model comparison results supported cloud cover and solar radiation as most important predictors, with *H5*$_{seasonal}$ (*Cloud* + *Srad* + *Tmin*) receiving the strongest support for the community and five out of twelve species, and *H4*$_{seasonal}$ (*Cloud* + *Srad*) receiving marginally more support than *H5*$_{seasonal}$ in two more species (Fig. 2). In all these cases, cloud cover showed a clear positive association with young leaf production (range $\beta_{Cloud}$ = 0.04-0.23; all 95% CrIs > 0; Fig. 2; Table S1). Solar radiation and minimum temperature were also positively associated with young leaf production when included (range $\beta_{Srad}$ =0.01-0.14; range $\beta_{Tmin}$ = 0.01-0.2; all 95% CrIs >0), except in two species where the 95% CrI overlapped with zero (*Celtis africana*: $\beta_{Srad}$=0.01, 95% CrI[-0.03,0.04], *Dombeya kirkii*: $\beta_{Srad}$=0.01, 95% CrI [-0.02,0.05], $\beta_{Tmin}$ =0.03, 95% CrI [-0.01,0.06]).

*H2*$_{seasonal}$ (*Rain* + *Tmin* + *Srad*) was best supported for one species (*Prunus africana*), where solar radiation had a positive effect ($\beta_{Srad}$=0.06, 95% CrI [0.03,0.1]), minimum temperature showed a negative effect ($\beta_{Tmin}$=-0.1, 95% CrI [-0.14, -0.07]), and rainfall no strong effect ($\beta_{Rain}$=-0.02, 95% CrI [-0.05,0.02]). *H3*$_{seasonal}$ (*Rain* + *Tmin*) was best supported for one species (*Parinari excelsa*), but with no clear associations ($\beta_{Rain}$=-0.02, 95% CrI [-0.05,0.02; $\beta_{Tmin}$=-0.03, 95% CrI [-0.14,0.01].

*H1*$_{seasonal}$ (*Srad* + *Rain*) received the strongest support in three species, with rainfall showing a positive association with young leaf production in all three species (range $\beta_{Rain}$ =0.05-0.13). However effect directions for solar radiation differed by species, with a negative association in one species ($\beta_{Srad}$=-0.05, 95% CrI [-0.09,-0.003]), a positive association in another species ($\beta_{Srad}$=0.04, 95% CrI[0.01,0.07]), and no clear association in the third species. The species without distinct seasonal peaks and lows (*Strombosia scheffleri*) was excluded from this seasonal climate analysis.

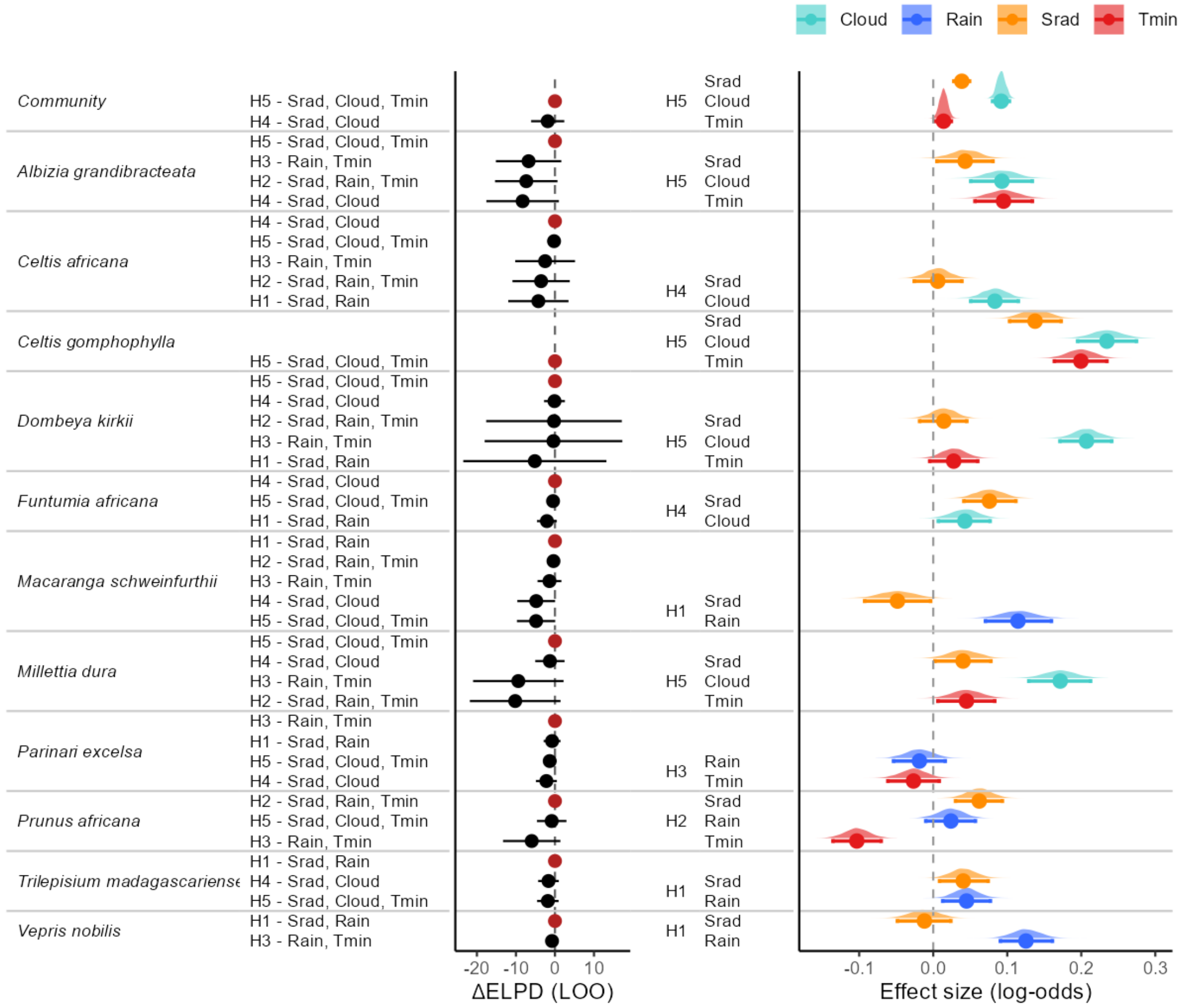

***Figure 2: Climate effects on seasonal young leaf production across species.*** *Model comparison results are shown in the left part with ΔELPD-LOO values for each hypothesis relative to the best-supported model for each species and for the community. Only models within 2 standard errors (SE) of the lowest ELPD-LOO difference are shown. Red points indicate the model with the lowest ELPD-LOO (i.e., the best predictive performance). Posterior estimates of climate predictors from the top-ranked seasonal model for each species and for the community are shown on the right. Points indicate posterior means, horizontal lines show 95% credible intervals, and shaded half-eye curves represent density of the posterior distributions of effect sizes. Estimates are shown on the log-odds scale, with the vertical dashed line indicating no effect, and positive/negative values indicating higher/lower leaf production with increasing values of the corresponding climate variable. Only climate predictors included in the top-ranked model for each species are shown. All estimates shown here are listed in Table S1 (SI).*

## Long-term patterns of young leaf production

The long-term patterns of young leaf production at the community level revealed some variability over the 26-year study period (Fig. 3). We identified one clearly distinct peak and one low (Fig.3, Table 3). The global peak occurred in August 2019 ($YL_{mean}$ = 2.54, 95% CrI [2.3, 2.77]), whereas the global low occurred in May 2000 ($YL_{mean}$ = 1.58, 95% CrI [1.39, 1.78]). The overall long-term amplitude at the community level *was* $\Delta YL_{mean}$ = 1.08.

The number of peaks, lows, and amplitude varied among species. Some species displayed multiple distinct peaks and lows distributed across the study, while others exhibited only a single identifiable peak and low. For the amplitude of young leaf production, eight species showed some long-term variation ($\Delta YL_{mean}$ between 1 and 2), while four species showed only little variation ($\Delta YL_{mean}$ between 0.6 and 1) (Fig. 3, Table 3).

The temporal structure was shared by most species, despite differences in the magnitude and number of peaks/lows. Five species experienced their global low between April 2005 and June 2006, and one additional species showed a clear low in August 2006. All but one species reached their highest peak between February 2016 and December 2019. More specifically, nine Species reached their highest peak in young leaf production between January 2018 and December 2019, and two species between February 2016 and February 2017 (Fig. 3, Table 3). The remaining species also reached a local peak in February 2017, but had an even higher peak in November 2022. Four species exhibited a subsequent low between April 2021 and January 2022.

***Table 3: Amplitude and timing of long-term peaks and lows in young leaf production for the community and the 12 different species.*** *Amplitude represents the difference ($\Delta YL_{mean}$) between the global maximum and minimum young leaf scores ($YL_{mean}$; original scale 0-6) predicted by the trend smooth term (TSp + intercept) derived from Bayesian generalized additive mixed models. For each peak and low, the Year-Month of occurrence and the respective $YL_{mean}$ value are given, with 95% credible intervals in brackets. Dashes indicate that no distinct long-term peak or low was identified for that species. Global peaks and lows are highlighted in bold.*

| Species | Amplitude | Peaks | Lows |
|---|---|---|---|
| *Community* | 1.10 | **2019-06: 2.54 [2.3, 2.77]** | 2000-05: 1.58 [1.39, 1.78] |
| *Albizia grandibracteata* | 1.04 | **2018-09: 2.34 [2.15, 2.53]** | **2005-04: 1.3 [1.15, 1.46]** |
| *Celtis africana* | 0.55 | **2018-12: 2.4 [2.22, 2.6]** | |
| *Celtis gomphophylla* | 0.92 | 2004-07: 1.93 [1.71, 2.18]<br>2010-02: 1.96 [1.76, 2.16]<br>2014-10: 1.8 [1.57, 2.05]<br>2017-02: 2.02 [1.82, 2.24]<br>**2022-11: 2.09 [1.89, 2.3]** | 2000-08: 1.42 [1.25, 1.61]<br>**2013-08: 1.25 [1.07, 1.44]**<br>2021-04: 1.53 [1.34, 1.72]<br>2024-01: 1.63 [1.42, 1.85] |
| *Dombeya kirkii* | 1.70 | 2002-01: 2.26 [2.04, 2.49]<br>2008-01: 2 [1.78, 2.23]<br>2010-07: 2.48 [2.27, 2.71]<br>**2019-03: 3.25 [3.03, 3.46]**<br>2024-04: 3.08 [2.84, 3.34] | **2005-09: 1.55 [1.36, 1.74]**<br>2021-05: 2.58 [2.33, 2.82] |
| *Funtumia africana* | 1.42 | 2008-09: 1.4 [1.23, 1.6]<br>2011-09: 1.68 [1.51, 1.89]<br>**2016-02: 2.24 [2.06, 2.42]** | **2006-06: 1 [0.86, 1.14]** |
| *Macaranga schweinfurthii* | 1.30 | 2003-11: 2.31 [2.1, 2.52]<br>2009-09: 2.3 [2.07, 2.55]<br>**2017-02: 2.52 [2.3, 2.75]** | 2006-08: 1.82 [1.62, 2.01]<br>**2014-04: 1.59 [1.25, 1.89]** |
| *Millettia dura* | 1.02 | 2009-12: 1.7 [1.47, 1.93]<br>**2018-02: 2.15 [1.9, 2.38]** | **2005-12: 1.12 [0.95, 1.31]**<br>2021-12: 1.46 [1.24, 1.68] |
| *Parinari excelsa* | 1.85 | 2008-01: 2.47 [2.28, 2.69]<br>2014-10: 2.97 [2.76, 3.19]<br>**2019-04: 3.43 [3.22, 3.66]** | |
| *Prunus africana* | 0.63 | **2019-01: 2.42 [2.22, 2.65]** | |
| *Strombosia scheffleri* | 1.24 | **2019-12: 2.23 [2.07, 2.38]** | **2006-02: 0.99 [0.89, 1.09]** |
| *Trilepisium madagascariense* | 0.93 | **2019-12: 2.62 [2.43, 2.84]** | **2009-05: 1.69 [1.54, 1.83]** |
| *Vepris nobilis* | 1.03 | **2018-01: 2.36 [2.18, 2.57]** | **2022-01: 1.96 [1.75, 2.14]** |

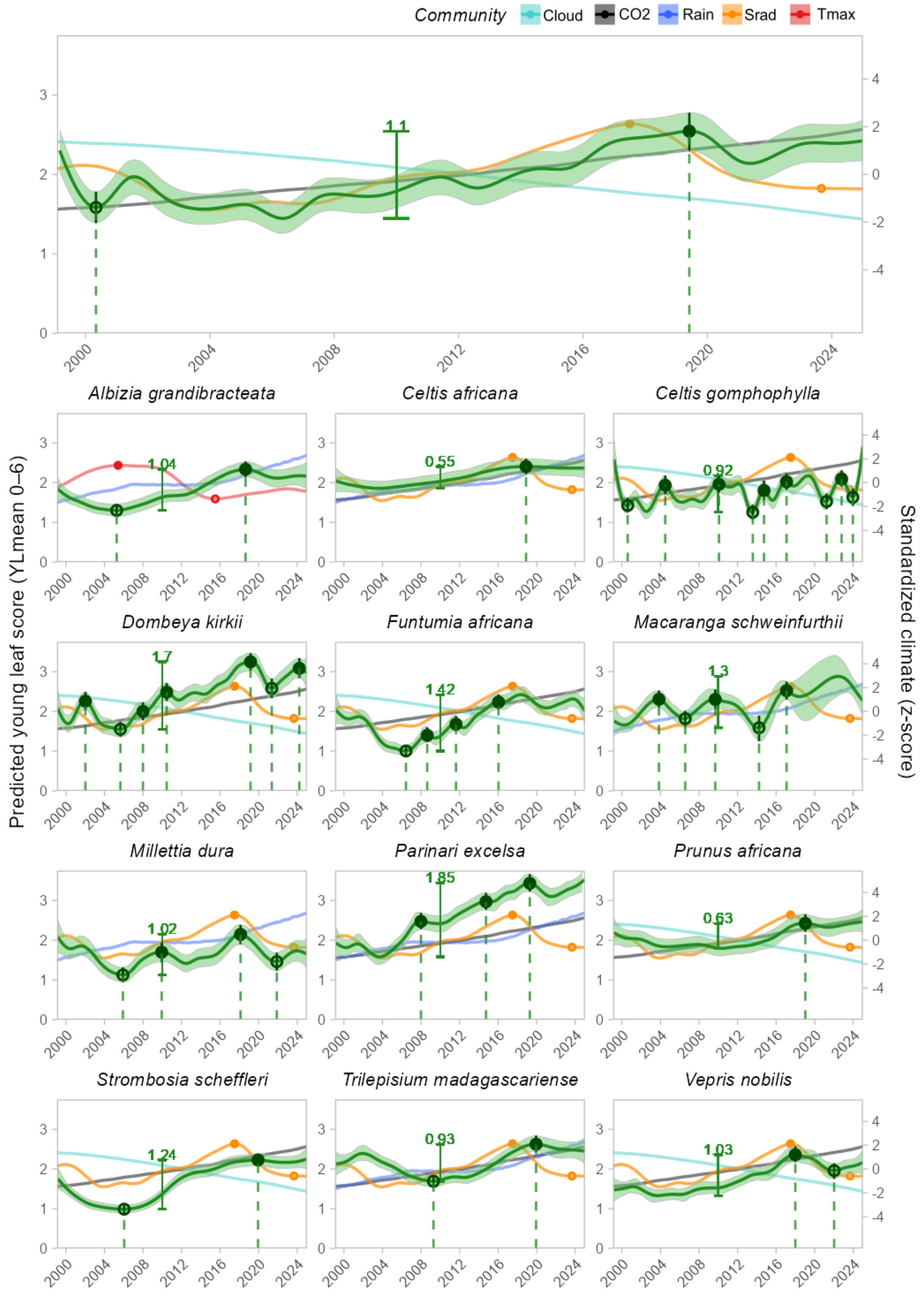

***Figure 3. Long-term patterns in young leaf production across the community and individual species.*** *Shown in green are predictions for young leaf scores ($YL_{mean}$; original scale 0-6) based on the trend smooth term (TSp + intercept) from the generalized additive mixed models (GAMM). Solid lines represent posterior means, shaded band indicate 95% credible intervals (CrI). Points mark the timing of clear long-term peaks and lows, with dashed vertical lines indicating their positions on the x-axis. Green, vertical segments illustrate the amplitude ($\Delta Y_{Lmean}$, difference between global peaks and lows). The trend smooth term (TSp) of the climate variables included in the most informative model for each species is presented in the respective species plot. For visual comparison across variables, climate variables were standardized (z-scaled) by subtracting the mean and dividing by the standard deviation. Only extrema with non-overlapping 95% credible intervals were marked with points.*

Association between long-term patterns of climatic variables and young leaf production

With regard to long-term trends, model comparison indicated a high degree of consistency across species (Fig. 4). For the community and 10 out of 12 species, long-term variation in young leaf production was best explained by two closely related hypotheses, $H3_{long\text{-}term}$ (*Srad* + $CO_2$ + *Cloud*) and $H4_{long\text{-}term}$ (*Srad* + $CO_2$ + *Rain*). For the community and nine species, the two models representing these hypotheses clearly outperformed all other models in terms of predictive performance, whereas in one species (*Trilepisium madagascariense*), $H4_{long\text{-}term}$ was the top-ranked model but differences to other models were smaller (Fig. 4).

Atmospheric CO2 was the strongest long-term predictor of young leaf production whenever included (range $\beta_{CO_2}$ = 0.05–0.38, 95% CrI >0; Fig. 4, Table S2). Solar radiation was also clearly positively associated with young leaf production at the community level and in nine out of twelve species (range $\beta_{Srad}$ = 0.05-0.17, 95% CrI >0), and in one species associated with young leaf production but with the 95% CrI intersecting with zero (*Celtis africana:* $\beta_{Srad}$ = 0.02, 95% CrI [-0.01, 0.06]). Cloud cover showed a positive association with young leaf production (range $\beta_{Cloud}$ = 0.03-0.09), except for two species where the 95% CrI intersected with zero. For the three species where rainfall was included, effects were more variable. One species exhibited a negative association with rainfall (*Celtis africana*: $\beta_{Rain}$ = -0.04, 95% CrI [-0.07, -0.01]), while rainfall showed no clear long-term association in the two others.

Three species deviated from this pattern. In two species (*Millettia dura, Macaranga schweinfurthii*), $H1_{long\text{-}term}$ (*Srad* + *Rain*) was the best supported hypothesis, with rainfall positively associated with young leaf production in both species (range $\beta_{Rain}$ = 0.04-0.05). However, while for *Millettia dura*, the model showed a positive association with solar radiation ($\beta_{Srad}$= 0.11, 95% CrI [0.08,0.15]), *Macaranga schweinfurthii*, was the only species exhibiting a negative relationship with solar radiation ($\beta_{Srad}$= -0.08, 95% CrI [-0.13,-0.03]). $H5_{long\text{-}term}$ (*Rain* + *Tmax*) best explained long-term variation in young leaf production only for *Albizia grandibracteata*, with maximum temperature showing a negative association with young leaf production ($\beta_{Tmax}$= -0.11, 95% CrI [-0.15,-0.08]), and rainfall a positive effect ($\beta_{Rain}$= 0.05, 95% CrI [0.02, 0.09]).

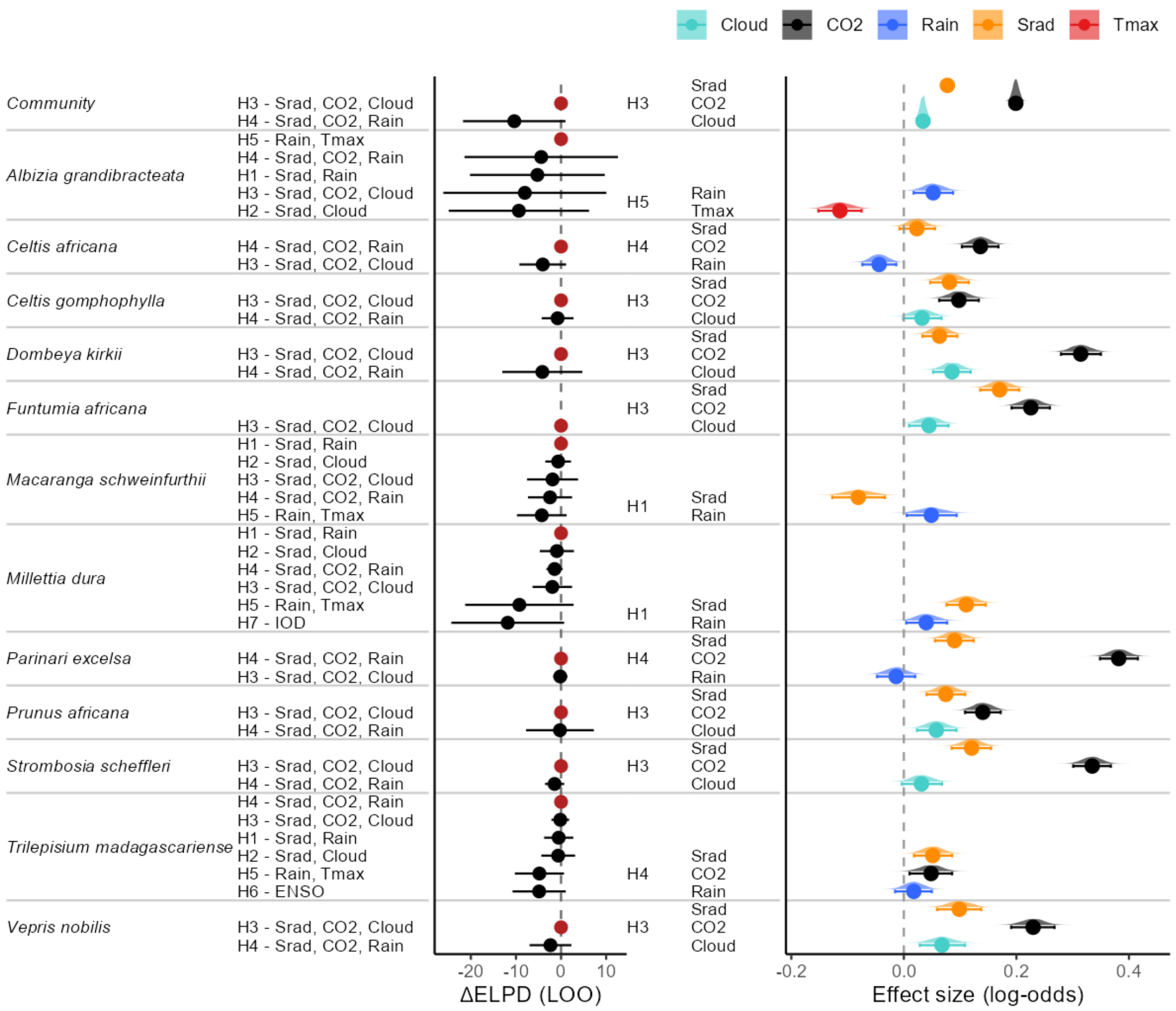

***Figure 4. Climate effects on long-term young leaf production across species**. Model comparison results are shown in the left part alongside effect estimates for the hypotheses as ΔELPD-LOO values relative to the best-supported model for each species and for the community. Only models within 2 standard errors (SE) of the lowest ELPD-LOO difference are shown. Red points indicate the model with the lowest ELPD-LOO (i.e., the best predictive performance). Posterior estimates of climate predictors from the top-ranked long-term model for each species and for the community are shown in the right part. Points indicate posterior means, horizontal lines show 95% credible intervals, and shaded half-eye curves represent the densities of posterior distributions of effect sizes. Estimates are shown on the log-odds scale, with the vertical dashed line indicating no effect, and positive/negative values indicating higher/lower leaf production with increasing values of the corresponding climate variable. Only climate predictors included in the top-ranked model for each species are shown. All estimates shown here are listed in Table S2 (SI).*

## Discussion

Young leaf production in Kibale exhibited consistent seasonality at the community level, with most species flushing during the two rainy seasons but differing in the timing and magnitude of their peaks. Over longer timescales, most species showed similar patterns with a global low between April 2005 and June 2006, and a global peak between January 2018 and December 2019, and multiple species showed a subsequent low between April 2021 and January 2022.

Across seasonal and long-term scales, young leaf production was most consistently associated with changes in light conditions, particularly solar radiation and cloud cover, suggesting that light is a primary driver of leaf phenological dynamics in this system. In addition, seasonal variability in leaf production was associated with rainfall and minimum temperature. Long-term variability was linked to rising atmospheric $CO_2$, rainfall, and maximum temperature. Together, this suggests that light availability was a driver of seasonal and long-term changes, whereas other climate variables may affect young leaf production differently depending on the temporal scale and species.

Seasonal trends in young leaf phenology and climatic drivers

Seasonal young leaf production exhibited one to two peaks for the community and most species but with inter-specific variation in timing and amplitude. Young leaf production was most consistently associated with solar radiation and cloud cover ($H4_{seasonal}$) and solar radiation and rainfall ($H1_{seasonal}$), and both combined with minimum temperature ($H5_{seasonal}$ and $H2_{seasonal}$). Seasonal leaf flushing has long been linked to increasing solar radiation when water is not a limiting factor (van Schaik et al., 1993; Wagner et al., 2017; Wright & van Schaik, 1994; Xiao et al., 2006) and water is likely not limiting in Kibale. Our results reinforce the central role of light availability in driving seasonal leaf flushing in moist tropical forests, while suggesting that cloud cover, likely through its promotion of diffuse radiation, plays an important role associated with young leaf production. Diffuse light can enhance canopy-scale photosynthesis and light-use efficiency, particularly under humid, cloudy conditions typical for rain seasons (Berry & Goldsmith, 2020; Roderick et al., 2001; van Diepen et al., 2025; Yan et al., 2017). This may hold especially true for understory and canopy level species, which comprise most of the species included in our analysis. For these species, rising solar radiation combined with increasing cloud cover at the onset of the rains may therefore provide an optimal window for leaf flushing, allowing new leaves to mature under favorable photosynthetic conditions.

By contrast, minimum temperature was not consistently associated with young leaf production, with species-specific responses being positive and negative. This suggests that temperature may modulate phenology in a more nuanced and context-dependent way (Jose et al., 2025; Kobayashi et al., 2020). Additionally, our findings emphasize that water availability and cloud dynamics remain important determinants of seasonal young leaf production, even where drought stress is limited, such as in Kibale (Chapman et al., 2021). Under projected climate change, however, regional increases in temperature and shifts in rainfall regimes (IPCC, 2023), together with potential declines in surface solar radiation and increases in cloud cover over East Africa (Adigun et al., 2025), may alter these seasonal light conditions. Because young leaf production in several species appears to track diffuse light availability or rainfall, reductions in solar radiation and increasing rainfall variability could dampen or alter the timing of leaf flushing events, potentially shifting both the magnitude and synchrony of young leaf availability across species.

Long-term trends in young leaf phenology and climatic drivers

Over the 26-years, young leaf production showed long-term variation with clear peaks and lows at the community and species levels and our results indicate that $CO_2$, solar radiation, and cloud cover act as important climate predictors for this long-term variation. Solar radiation and atmospheric $CO_2$ combined with either cloud cover or rainfall ($H3_{long\text{-}term}$ and $H4_{long\text{-}term}$) emerged as the best hypotheses for the majority of species. Interestingly, we found that large-scale climate phenomena such as ENSO ($H6_{long\text{-}term}$) and IOD ($H7_{long\text{-}term}$) predicted long-term phenology patterns in Kibale less than expected from results for other phenophases, such as fruit, at the same field site (Chapman et al., 2018). Despite recent evidence suggesting a weakening of the $CO_2$ fertilization effect from 1982 to 2015 (Wang et al., 2020), the strong positive association between atmospheric $CO_2$ and young leaf production in our study supports an ongoing greening trend consistent with global observations (Piao et al., 2020). However, low levels of young leaf production around 2006 and a decline in young leaf production around 2021 after the strong peak in 2018/19 indicate that $CO_2$ alone cannot explain long-term phenological dynamics, as $CO_2$ increased steadily over the study.

Long-term variation in young leaf production in Kibale also tracked changes in surface solar radiation. This suggests that light availability can act as a constraint modulating potential $CO_2$-driven gains in leaf production, which is in line with experimental evidence that light availability can limit $CO_2$ uptake in tropical tree species during rainy conditions (Graham et al., 2003). Interestingly, solar radiation exhibited a low from 2003 to 2007, then increased until it reached a peak in 2017, and then declined in the following years (SI Fig. S2). These long-term changes appear to be influenced by variation in solar activity, which follows an approximately 11-year cycle (Hathaway, 2015; Hempelmann & Weber, 2012) and may therefore affect leaf production. If projected declines in solar radiation and increasing cloud cover in Uganda materialize (Adigun et al., 2025), light limitation may become more pronounced in the future, potentially constraining $CO_2$-associated gains in leaf production despite continued atmospheric enrichment.

These dynamics can have cascading effects on specialist folivores. The young leaves of just four tree species alone (*Prunus africana*, *Celtis africana*, *Celtis gomphophylla*, and *Trilepisium madagascariense)* account for approximately 28% of the diet of a red colobus group studied in Kibale (Lauer et al., 2025). Young leaves provide higher protein content and lower fiber than mature foliage and are therefore disproportionately important for colobus nutrition (Chapman & Chapman, 2002). Even modest changes in the availability, timing, duration, or synchrony of leaf flushing in these species could alter periods of peak food availability, increase foraging effort, or reduce dietary quality.

Importantly, our results indicate that these key food species do not respond uniformly to climatic variation. Instead, they exhibit distinct, and in some cases opposing, sensitivities to temperature, solar radiation, and rainfall. For example, while increases in minimum temperature were positively associated with young leaf production in *Celtis gomphophylla*, it was negatively associated in *Prunus africana*. With temperatures projected to increase (IPCC, 2023), such contrasting responses are likely to reshape the seasonal structure and reliability of preferred resources. Similarly, although rising atmospheric $CO_2$ was associated with increased young leaf production across species at longer time scales, positive associations with solar radiation in three of the four species suggest that projected regional declines in radiation could constrain or offset $CO_2$-driven increases. Together, these species-specific patterns indicate that climate change will not affect all food species equally, increasing temporal variability and affecting resource composition rather than uniformly enhancing resource abundance. For an endangered and relatively specialized primate, such changes could influence reproductive success, juvenile survival, and ultimately population trajectories.

Our results show that young leaf production is shaped by interacting climatic drivers across temporal scales, with diffuse light conditions (solar radiation and cloud cover) and atmospheric $CO_2$ playing central roles. Because these drivers are projected to shift under ongoing climate change (Adigun et al., 2025; IPCC, 2023), phenological dynamics in tropical forests are likely to be reshaped in complex and species-specific ways. Community-level patterns may therefore mask biologically meaningful differences among key food species, particularly for specialist herbivores, such as red colobus (Lauer et al., 2025).

Greater understanding of how specific functional or taxonomic groups respond to projected climatic shifts could therefore allow us to anticipate changes in the availability of key food resources and forecast population trajectories for specialized and endangered consumers, such as red colobus. This knowledge has direct conservation implications. Current restoration efforts in Kibale prioritize planting species that support primate population growth, yet species selection does not explicitly account for how those trees may perform under future climate conditions. Integrating phenological sensitivity to climate change into restoration planning could improve the resilience of regenerating forests and help secure the long-term food availability for consumers and sustain resilient, functioning tropical forest ecosystems.

**Conclusion**

Our study demonstrates that young leaf phenology in a moist tropical forest exhibits seasonal and long-term variation, with likely implications for plants and their consumers. Seasonally, solar radiation, cloud cover, rainfall, and minimum temperature best predicted young leaf production, while over longer timescales, atmospheric $CO_2$, cloud cover, and solar radiation - linked to solar cycles - emerged as the best predictors. Our results confirm that the $CO_2$ fertilization effect appears to contribute to long-term increases in young leaf production, but declines at the beginning and end of the study suggest a constraining role of solar radiation.

Importantly, we detected substantial inter-specific variation in both seasonal and long-term responses, emphasizing that community-level patterns can mask biologically important species-specific dynamics and drivers. These findings highlight the need to consider multiple climatic drivers when evaluating long-term phenological change and long-term datasets. More broadly, we confirm the important role of increasing $CO_2$ levels on leaf flushing and suggest that solar radiation cycles may be an overlooked mechanism shaping long-term tropical forest productivity, with direct implications for understanding ecosystem dynamics and shaping conservation strategies under global change.

## Acknowledgments

We thank all the people who made the collection of the analysed long-term data possible, specifically the project managers and field assistants of the Kibale Fish and Monkey Project, including Dennis Twinomugisha, Emmanuel Opito, Martin Golooba, Emmanuel Aliganyira, Robert Basaija, Clovis Kaganzi, Peter Tuhairwe, and particularly Tusiime Laurence. Furthermore, we thank the many people who, over the last 20+ years, made major contribution to entering and cleaning the data set. We thank the Baden-Württemberg Stiftung for their support of the project "The cascading effects of climate change on primate food availability, behaviour, and survival" to Urs Kalbitzer. Furthermore, Urs Kalbitzer was supported by funding from the Center for the Advanced Study of Collective Behavior (CASCB) and the Young Scholar Fund (YSF) at the University of Konstanz, Germany. Additionally, Laura Lüthy was supported by the International Max Planck Research School for Quantitative Behaviour, Ecology and Evolution (IMPRS-QBEE). Funding for the long-term behavioural data collection in Kibale was provided by many sources, but primarily by the Canada Research Chairs Program, Natural Science and Engineering Research Council of Canada, Fonds Québécois de la Recherché sur la Nature et les Technologies, the National Geographic Society, Wildlife Conservation Society, and the Max-Planck Institute of Animal Behavior.

**Author Contributions:** L.L. C.C., and U.K. conceived the study; L.L. and U.K. analyzed the data; L.L. drafted the manuscript; U.K., C.C., P.L. and P.O. provided scientific input and reviewed the manuscript.

## References

Abatzoglou, J. T., Dobrowski, S. Z., Parks, S. A., & Hegewisch, K. C. (2018). TerraClimate, a high-resolution global dataset of monthly climate and climatic water balance from 1958–2015. *Scientific Data 2018 5:1*, *5*(1), 1–12. https://doi.org/10.1038/sdata.2017.191

Abernethy, K., Bush, E. R., Forget, P. M., Mendoza, I., & Morellato, L. P. C. (2018). Current issues in tropical phenology: A synthesis. *Biotropica*, *50*(3), 477–482. https://doi.org/10.1111/btp.12558

Adigun, P., Ogunrinde, A. T., Dairaku, K., Adebiyi, A. A., & Xian, X. (2025). The future of photovoltaic energy potential in Africa under higher emission scenarios: Insights from CMIP6 multi-model ensemble analysis. *Solar Energy*, *285*, 113078. https://doi.org/10.1016/j.solener.2024.113078

Ainsworth, E. A., & Rogers, A. (2007). The response of photosynthesis and stomatal conductance to rising [$CO_2$]: Mechanisms and environmental interactions. *Plant, Cell & Environment*, *30*(3), 258–270. https://doi.org/10.1111/j.1365-3040.2007.01641.x

Barlow, J., França, F., Gardner, T. A., Hicks, C. C., Lennox, G. D., Berenguer, E., Castello, L., Economo, E. P., Ferreira, J., Guénard, B., Gontijo Leal, C., Isaac, V., Lees, A. C., Parr, C. L., Wilson, S. K., Young, P. J., & Graham, N. A. J. (2018). The future of hyperdiverse tropical ecosystems. *Nature*, *559*(7715), 517–526. https://doi.org/10.1038/s41586-018-0301-1

Barone, J. A. (1998). Effects of light availability and rainfall on leaf production in a moist tropical forest in central Panama. *Journal of Tropical Ecology*, *14*, 309–321. https://doi.org/10.1017/S0266467498000248

Berry, Z. C., & Goldsmith, G. R. (2020). Diffuse light and wetting differentially affect tropical tree leaf photosynthesis. *New Phytologist*, *225*(1), 143–153. https://doi.org/10.1111/nph.16121

Bürkner, P.-C. (2017). **brms**: An *R* Package for Bayesian Multilevel Models Using *Stan*. *Journal of Statistical Software*, *80*(1), 1–28. https://doi.org/10.18637/jss.v080.i01

Chapman, C. A., & Chapman, L. J. (2002). Foraging challenges of red colobus monkeys: Influence of nutrients and secondary compounds. *Comparative Biochemistry and Physiology Part A: Molecular & Integrative Physiology*, *133*(3), 861–875. https://doi.org/10.1016/S1095-6433(02)00209-X

Chapman, C. A., Chapman, L. J., Jacob, A. L., Rothman, J. M., Omeja, P., Reyna-Hurtado, R., Hartter, J., & Lawes, M. J. (2010). Tropical tree community shifts: Implications for wildlife conservation. *Biological Conservation*, *143*(2), 366–374. https://doi.org/10.1016/j.biocon.2009.10.023

Chapman, C. A., Galán-Acedo, C., Gogarten, J. F., Hou, R., Lawes, M. J., Omeja, P. A., Sarkar, D., Sugiyama, A., & Kalbitzer, U. (2021). A 40-year evaluation of drivers of African rainforest change. *Forest Ecosystems*, *8*(1). https://doi.org/10.1186/s40663-021-00343-7

Chapman, C. A., & Lambert, J. E. (2000). Habitat alteration and the conservation of African primates: Case study of Kibale National Park, Uganda. *American Journal of Primatology*, *50*(3), 169–185. https://doi.org/10.1002/(SICI)1098-2345(200003)50:3%253C169::AID-AJP1%253E3.0.CO;2-P

Chapman, C. A., Valenta, K., Bonnell, T. R., Brown, K. A., & Chapman, L. J. (2018). Solar radiation and ENSO predict fruiting phenology patterns in a 15-year record from Kibale National Park, Uganda. *Biotropica*, *50*(3), 384–395. https://doi.org/10.1111/btp.12559

Coley, P. D., & Barone, J. A. (1996). Herbivory and Plant Defenses in Tropical Forests. *Annual Review of Ecology and Systematics*, *27*, 305–335.

Davis, C. C., Lyra, G. M., Park, D. S., Asprino, R., Maruyama, R., Torquato, D., Cook, B. I., & Ellison, A. M. (2022). New directions in tropical phenology. *Trends in Ecology & Evolution*, *37*(8), 683–693. https://doi.org/10.1016/j.tree.2022.05.001

Durand, M., Murchie, E. H., Lindfors, A. V., Urban, O., Aphalo, P. J., & Robson, T. M. (2021). Diffuse solar radiation and canopy photosynthesis in a changing environment. *Agricultural and Forest Meteorology*, *311*, 108684. https://doi.org/10.1016/j.agrformet.2021.108684

Gorelick, N., Hancher, M., Dixon, M., Ilyushchenko, S., Thau, D., & Moore, R. (2017). Google Earth Engine: Planetary-scale geospatial analysis for everyone. *Remote Sensing of Environment*, *202*, 18–27. https://doi.org/10.1016/J.RSE.2017.06.031

Graham, E. A., Mulkey, S. S., Kitajima, K., Phillips, N. G., & Wright, S. J. (2003). Cloud cover limits net $CO_2$ uptake and growth of a rainforest tree during tropical rainy seasons. *Proceedings of the National Academy of Sciences*, *100*(2), 572–576. https://doi.org/10.1073/pnas.0133045100

Gu, L., Baldocchi, D., Verma, S. B., Black, T. A., Vesala, T., Falge, E. M., & Dowty, P. R. (2002). Advantages of diffuse radiation for terrestrial ecosystem productivity. *Journal of Geophysical Research: Atmospheres*, *107*(D6), ACL 2-1-ACL 2-23. https://doi.org/10.1029/2001JD001242

Gulev, S. K., Thorne, P. W., Ahn, J., Dentener, F. J., Domingues, C. M., Gerland, S., Gong, D., Kaufman, D. S., Nnamchi, H. C., & Quaas, J. (2021). Changing state of the climate system. *Climate Change 2021: The Physical Science Basis. Contribution of Working Group I to the Sixth Assessment Report of the Intergovernmental Panel on Climate Change*, 287–422.

Harris, I., Osborn, T. J., Jones, P., & Lister, D. (2020). Version 4 of the CRU TS monthly high-resolution gridded multivariate climate dataset. *Scientific Data*, *7*(1), 109. https://doi.org/10.1038/s41597-020-0453-3

Hathaway, D. H. (2015). The Solar Cycle. *Living Reviews in Solar Physics*, *12*(1), 4. https://doi.org/10.1007/lrsp-2015-4

Hempelmann, A., & Weber, W. (2012). Correlation Between the Sunspot Number, the Total Solar Irradiance, and the Terrestrial Insolation. *Solar Physics*, *277*(2), 417–430. https://doi.org/10.1007/s11207-011-9905-4

IPCC. (2023). *Climate Change 2022 – Impacts, Adaptation and Vulnerability: Working Group II Contribution to the Sixth Assessment Report of the Intergovernmental Panel on Climate Change* (1st ed.). Cambridge University Press. https://doi.org/10.1017/9781009325844

Jin, J., Jian, D., Zhou, X., Chen, Q., & Li, Y. (2025). Impact of El Niño–Southern Oscillation on Global Vegetation. *Atmosphere*, *16*(6). https://doi.org/10.3390/atmos16060701

Jose, K., Najeeb, N., Bandopadhyay, A., Singh, C. P., & Chaturvedi, R. K. (2025). Unraveling meteorological drivers of leaf phenology in the Western Ghats, India. *Trees, Forests and People*, *20*, 100861. https://doi.org/10.1016/j.tfp.2025.100861

Karlsson, K.-G., Riihelä, A., Trentmann, J., Stengel, M., Solodovnik, I., Meirink, J. F., Devasthale, A., Jääskeläinen, E., Kallio-Myers, V., Eliasson, S., Benas, N., Johansson, E., Stein, D., Finkensieper, S., Håkansson, N., Akkermans, T., Clerbaux, N., Selbach, N., Marc, S., & Hollmann, R. (2023). *CLARA-A3: CM SAF cLoud, Albedo and surface RAdiation dataset from AVHRR data—Edition 3*. Satellite Application Facility on Climate Monitoring (CM SAF). https://doi.org/10.5676/EUM_SAF_CM/CLARA_AVHRR/V003

Kobayashi, M. J., Ng, K. K. S., Lee, S. L., Muhammad, N., & Tani, N. (2020). Temperature is a regulator of leaf production in the family Dipterocarpaceae of equatorial Southeast Asia. *American Journal of Botany*, *107*(11), 1491–1503. https://doi.org/10.1002/ajb2.1557

Kobayashi, S., Ota, Y., Harada, Y., Ebita, A., Moriya, M., Onoda, H., Onogi, K., Kamahori, H., Kobayashi, C., Endo, H., Miyaoka, K., & Kiyotoshi, T. (2015). The JRA-55 reanalysis: General specifications and basic characteristics. *Journal of the Meteorological Society of Japan*, *93*(1), 5–48. https://doi.org/10.2151/jmsj.2015-001

Lan, X., Trans, P., & Thoning, K. W. (2024, July). *Trends in globally-averaged CO2 determined from NOAA Global Monitoring Laboratory measure-ments. Version 2024-07*. https://doi.org/10.15138/9N0H-ZH07

Lauer, P., Chapman, C. A., Omeja, P., Rothman, J. M., & Kalbitzer, U. (2025). A long-term study on food choices and nutritional goals of a leaf-eating primate. *Ecosphere*, *16*(1). https://doi.org/10.1002/ecs2.70162

Lewis, S. L., Malhi, Y., & Phillips, O. L. (2004). Fingerprinting the impacts of global change on tropical forests. *Philosophical Transactions of the Royal Society B: Biological Sciences*, *359*(1443), 437–462. https://doi.org/10.1098/rstb.2003.1432

Luse Belanganayi, B., Phaka, C. M., Djiofack, B. Y., Laurent, F., Liévens, K., Luambua, N. K., Bolaya, T., Bourland, N., Hubau, W., Beeckman, H., & De Mil, T. (2025). Timing of cambial phenology of rainforest trees as indicator of climate sensitivity of the Congo Basin biome. *Global Ecology and Conservation*, *62*, e03740. https://doi.org/10.1016/j.gecco.2025.e03740

Marchant, R., Mumbi, C., Behera, S., & Yamagata, T. (2007). The Indian Ocean dipole—The unsung driver of climatic variability in East Africa. *African Journal of Ecology*, *45*(1), 4–16. https://doi.org/10.1111/j.1365-2028.2006.00707.x

Matsuda, I., Ihobe, H., Tashiro, Y., Yumoto, T., Baranga, D., & Hashimoto, C. (2020). The diet and feeding behavior of the black-and-white colobus (Colobus guereza) in the Kalinzu Forest, Uganda. *Primates*, *61*(3), 473–484. https://doi.org/10.1007/s10329-020-00794-6

McElreath, R. (2018). *Statistical rethinking: A Bayesian course with examples in R and Stan*. Chapman and Hall/CRC. https://api.taylorfrancis.com/content/books/mono/download?identifierName=doi&identifierValue=10.1201/9781315372495&type=googlepdf

OpenAI. (2025). *ChatGPT* [Https://chat.openai.com/]. https://chat.openai.com/

Pearson, K. (1895). VII. Note on regression and inheritance in the case of two parents. *Proceedings of the Royal Society of London*, *58*(347–352), 240–242. https://doi.org/10.1098/rspl.1895.0041

Piao, S., Liu, Q., Chen, A., Janssens, I. A., Fu, Y., Dai, J., Liu, L., Lian, X., Shen, M., & Zhu, X. (2019). Plant phenology and global climate change: Current progresses and challenges. *Global Change Biology*, *25*(6), 1922–1940. https://doi.org/10.1111/gcb.14619

Piao, S., Wang, X., Park, T., Chen, C., Lian, X., He, Y., Bjerke, J. W., Chen, A., Ciais, P., Tømmervik, H., Nemani, R. R., & Myneni, R. B. (2020). Characteristics, drivers and feedbacks of global greening. *Nature Reviews Earth & Environment*, *1*(1), 14–27. https://doi.org/10.1038/s43017-019-0001-x

Polansky, L., & Robbins, M. M. (2013). Generalized additive mixed models for disentangling long-term trends, local anomalies, and seasonality in fruit tree phenology. *Ecology and Evolution*, *3*(9), 3141–3151. https://doi.org/10.1002/ece3.707

Roderick, M. L., Farquhar, G. D., Berry, S. L., & Noble, I. R. (2001). On the direct effect of clouds and atmospheric particles on the productivity and structure of vegetation. *Oecologia*, *129*(1), 21–30. https://doi.org/10.1007/s004420100760

Rojo, J., Rivero, R., Romero-Morte, J., Fernández-González, F., & Pérez-Badia, R. (2017). Modeling pollen time series using seasonal-trend decomposition procedure based on LOESS smoothing. *International Journal of Biometeorology*, *61*(2), 335–348. https://doi.org/10.1007/s00484-016-1215-y

Rosenzweig, C. ;, Casassa, G. ;, Karoly, Imeson, J. ;, Liu, A. ;, Menzel, C. ;, Rawlins, A. ;, Root, S. ;, Seguin, T. L. ;, & Tryjanowski, B. ; (2007). *Assessment of observed changes and responses in natural and managed systems*. https://doi.org/10.5167/uzh-33180

Saji, N., & Yamagata, T. (2003). Possible impacts of Indian Ocean Dipole mode events on global climate. *Climate Research*, *25*, 151–169. https://doi.org/10.3354/cr025151

Simpson, G. L. (2018). Modelling Palaeoecological Time Series Using Generalised Additive Models. *Frontiers in Ecology and Evolution*, *6*. https://doi.org/10.3389/fevo.2018.00149

Stan Development Team. (2023). *Stan Reference Manual* [Computer software]. https://mc-stan.org.

Struhsaker, T. T. (1997). Ecology of an African rain forest: Logging in Kibale and the conflict between conservation and exploitation. *University of Florida Press, Gainesville*. https://www.cabidigitallibrary.org/doi/full/10.5555/20000617542

Sullivan, M. K., Fayolle, A., Bush, E., Ofosu-Bamfo, B., Vleminckx, J., Metz, M. R., & Queenborough, S. A. (2024). Cascading effects of climate change: New advances in drivers and shifts of tropical reproductive phenology. *Plant Ecology*, *225*(3), 175–187. https://doi.org/10.1007/s11258-023-01377-3

van Diepen, K. H. H., Kaiser, E., Hartogensis, O. K., Graf, A., de Arellano, J. V. G., & Moene, A. F. (2025). When do clouds and aerosols lead to higher canopy photosynthesis? *Agricultural and Forest Meteorology*, *370*. https://doi.org/10.1016/j.agrformet.2025.110597

van Schaik, C. P., Terborgh, J. W., & Wright, S. J. (1993). *The Phenology of Tropical Forests: Adaptive Significance and Consequences for Primary Consumers* (Vol. 24, pp. 353–377). https://about.jstor.org/terms

Vehtari, A., Gelman, A., & Gabry, J. (2017). Practical Bayesian model evaluation using leave-one-out cross-validation and WAIC. *Statistics and Computing*, *27*(5), 1413–1432. https://doi.org/10.1007/s11222-016-9696-4

Vehtari, A., Gelman, A., Simpson, D., Carpenter, B., & Bürkner, P.-C. (2021). Rank-Normalization, Folding, and Localization: An Improved Rˆ for Assessing Convergence of MCMC (with Discussion). *Bayesian Analysis*, *16*(2), 667–718. https://doi.org/10.1214/20-BA1221

Verbesselt, J., Hyndman, R., Newnham, G., & Culvenor, D. (2010). Detecting trend and seasonal changes in satellite image time series. *Remote Sensing of Environment*, *114*(1), 106–115. https://doi.org/10.1016/j.rse.2009.08.014

Wagner, F. H., Hérault, B., Rossi, V., Hilker, T., Maeda, E. E., Sanchez, A., Lyapustin, A. I., Galvão, L. S., Wang, Y., & Aragão, L. E. O. C. (2017). Climate drivers of the Amazon forest greening. *PLoS ONE*, *12*(7). https://doi.org/10.1371/journal.pone.0180932

Wang, S., Zhang, Y., Ju, W., Chen, J. M., Ciais, P., Cescatti, A., Sardans, J., Janssens, I. A., Wu, M., Berry, J. A., Campbell, E., Fernández-Martínez, M., Alkama, R., Sitch, S., Friedlingstein, P., Smith, W. K., Yuan, W., He, W., Lombardozzi, D., … Peñuelas, J. (2020). Recent global decline of $CO_2$ fertilization effects on vegetation photosynthesis. *Science*, *370*(6522), 1295–1300. https://doi.org/10.1126/science.abb7772

Wild, M., Roesch, A., & Ammann, C. (2012). Global dimming and brightening—Evidence and agricultural implications. *CABI Reviews*, 1–7. https://doi.org/10.1079/PAVSNNR20127003

Williams, I. N., Riley, W. J., Kueppers, L. M., Biraud, S. C., & Torn, M. S. (2016). Separating the effects of phenology and diffuse radiation on gross primary productivity in winter wheat. *Journal of Geophysical Research: Biogeosciences*, *121*(7), 1903–1915. https://doi.org/10.1002/2015JG003317

Williams, R. J., Myers, B. A., Muller, W. J., Duff, G. A., & Eamus, D. (1997). Leaf Phenology of Woody Species in a North Australian Tropical Savanna. *Ecology*, *78*(8), 2542–2558. https://doi.org/10.1890/0012-9658(1997)078%255B2542:LPOWSI%255D2.0.CO;2

Wright, S. J., & van Schaik, C. P. (1994). Light and the Phenology of Tropical Trees. *American Naturalist*, *143*(1), 192–199.

Xiao, X., Hagen, S., Zhang, Q., Keller, M., & Moore, B. (2006). Detecting leaf phenology of seasonally moist tropical forests in South America with multi-temporal MODIS images. *Remote Sensing of Environment*, *103*(4), 465–473. https://doi.org/10.1016/j.rse.2006.04.013

Yan, H., Wang, S. Q., da Rocha, H. R., Rap, A., Bonal, D., Butt, N., Coupe, N. R., & Shugart, H. H. (2017). Simulation of the Unexpected Photosynthetic Seasonality in Amazonian Evergreen Forests by Using an Improved Diffuse Fraction-Based Light Use Efficiency Model. *Journal of Geophysical Research: Biogeosciences*, *122*(11), 3014–3030. https://doi.org/10.1002/2017JG004008

Yang, X., Wu, J., Chen, X., Ciais, P., Maignan, F., Yuan, W., Piao, S., Yang, S., Gong, F., Su, Y., Dai, Y., Liu, L., Zhang, H., Bonal, D., Liu, H., Chen, G., Lu, H., Wu, S., Fan, L., … Wright, S. J. (2021). A comprehensive framework for seasonal controls of leaf abscission and productivity in evergreen broadleaved tropical and subtropical forests. *The Innovation*, *2*(4), 100154. https://doi.org/10.1016/j.xinn.2021.100154

Zhu, Z., Piao, S., Myneni, R. B., Huang, M., Zeng, Z., Canadell, J. G., Ciais, P., Sitch, S., Friedlingstein, P., Arneth, A., Cao, C., Cheng, L., Kato, E., Koven, C., Li, Y., Lian, X., Liu, Y., Liu, R., Mao, J., … Zeng, N. (2016). Greening of the Earth and its drivers. *Nature Climate Change*, *6*(8), 791–795. https://doi.org/10.1038/nclimate3004

# Supporting Information

<u>Patterns of climatic variables</u>

The rainfall model detected a bimodal seasonal pattern (Fig. 1a), with a smaller peak in April (170 mm; small rain season) and a larger peak in October (252 mm; large rain season), and rainfall minima in February (53 mm) and July (69 mm). The fraction of cloud cover followed a similar pattern (Fig. 1b), peaking in April (83 %) and October (85 %), and lows in January (65 %) and June (66 %). Maximum and minimum temperature only showed single main peaks, $T_{max}$ in February (28.8 °C, Fig. 1c), and $T_{min}$ later in April (16.7°C, Fig. 1d). The lowest $T_{max}$ (26.8 °C) and $T_{min}$ (15.5 °C) both occurred in July between the two rain seasons. Atmospheric $CO_2$ shows slight seasonality (Fig. 1e), with one peak in April (392.5 ppm) and a low in August (387.7 ppm). In contrast to other variables, solar radiation had three peaks per year (Fig. 1f), namely in February (205.1 $W/m^2$), May (200.0 $W/m^2$), and September (195.5 $W/m^2$), and was at its lowest in July (180.4 $W/m^2$). The MEI (indicating ENSO patterns, Fig. 1g) and DMI (indicating IOD patterns, Fig. 1h) showed no seasonality.

Over the 26-year study (1999–2024), the average model-predicted rainfall based on the trend spline remained relatively stable (1999: 59.2 mm; 2024: 64.0 mm, Fig. 2a). In contrast, cloud cover showed a slight decrease (1999: 67.8%; 2024: 63.7%, Fig. 2b). Both average yearly $T_{max}$ (1999: 26.8°C; 2024: 26.7°C, Fig. 2c) and $T_{min}$ (1999: 15.1°C; 2024: 15.1°C, Fig. 2d) showed no linear trend over the study, but there was a warmer period between 2000 and 2010. Global atmospheric $CO_2$ steadily increased from 367.5 ppm in 1999 to 422.4 ppm in 2024 (Fig. 2e). The solar radiation trend spline exhibited more pronounced variability, with a decline from January 1999 to a low in April 2004 (174.3 $W/m^2$), followed by an increase to a maximum in January 2017 (197.3 $W/m^2$), and another decline through the end of the study (Fig. 2f). The trend spline of the MEI shows several ENSO events during the study, with the most pronounced one in 2016 (Fig. 2g), while the trend spline of the DMI (IOD) exhibited less variability and the highest peak in 2019 (Fig. 2h)

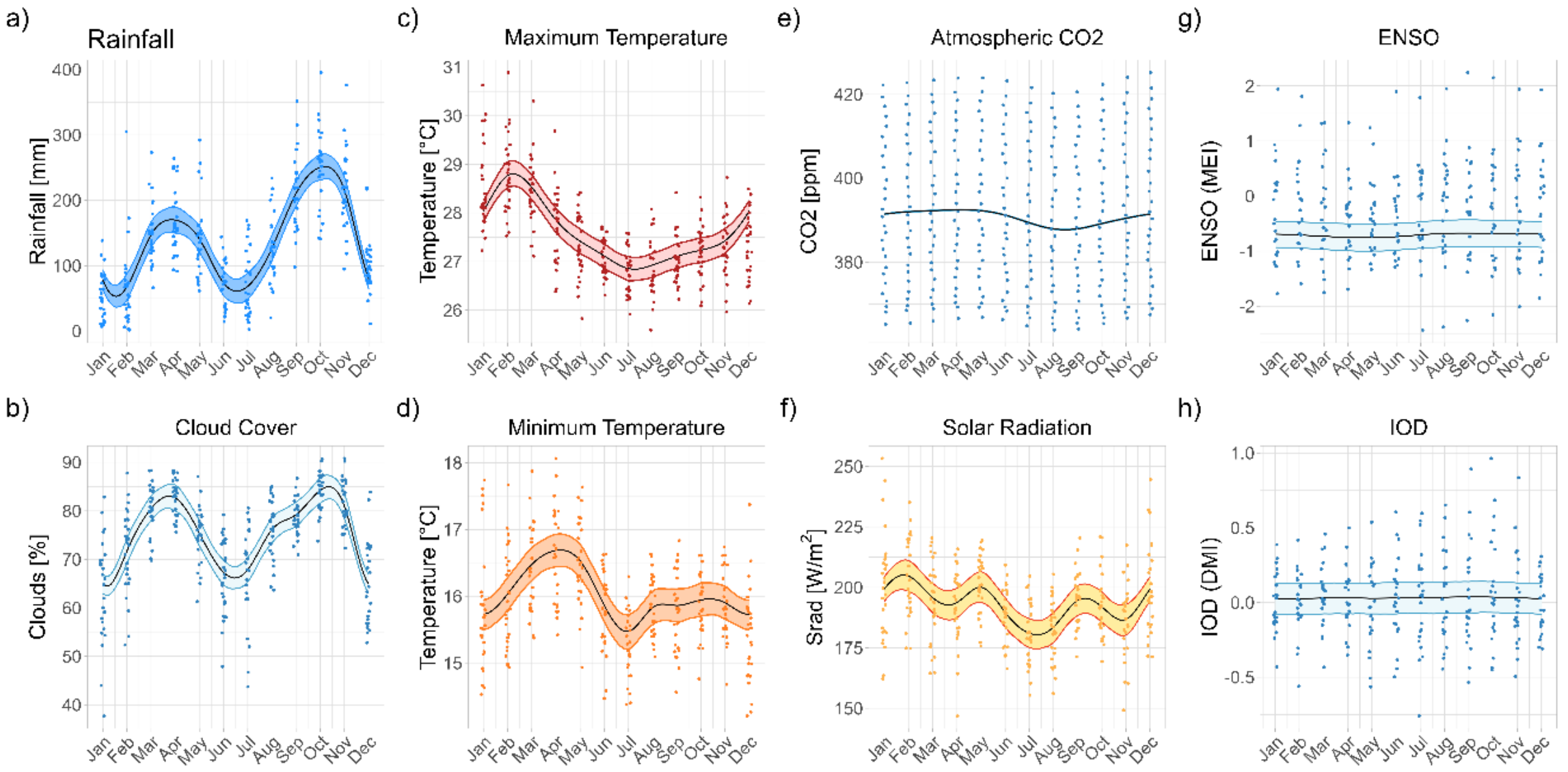

***Figure S1*** *Seasonal splines of the climate models for (a) rainfall, (b) fractional cloud cover, (c) Tmax, (d) Tmin, (e) atmospheric CO2, (f) solar radiation, (g) ENSO, and (h) IOD. Shaded areas represent the 95% credible interval of the estimated mean. Jittered points show the observed data.*

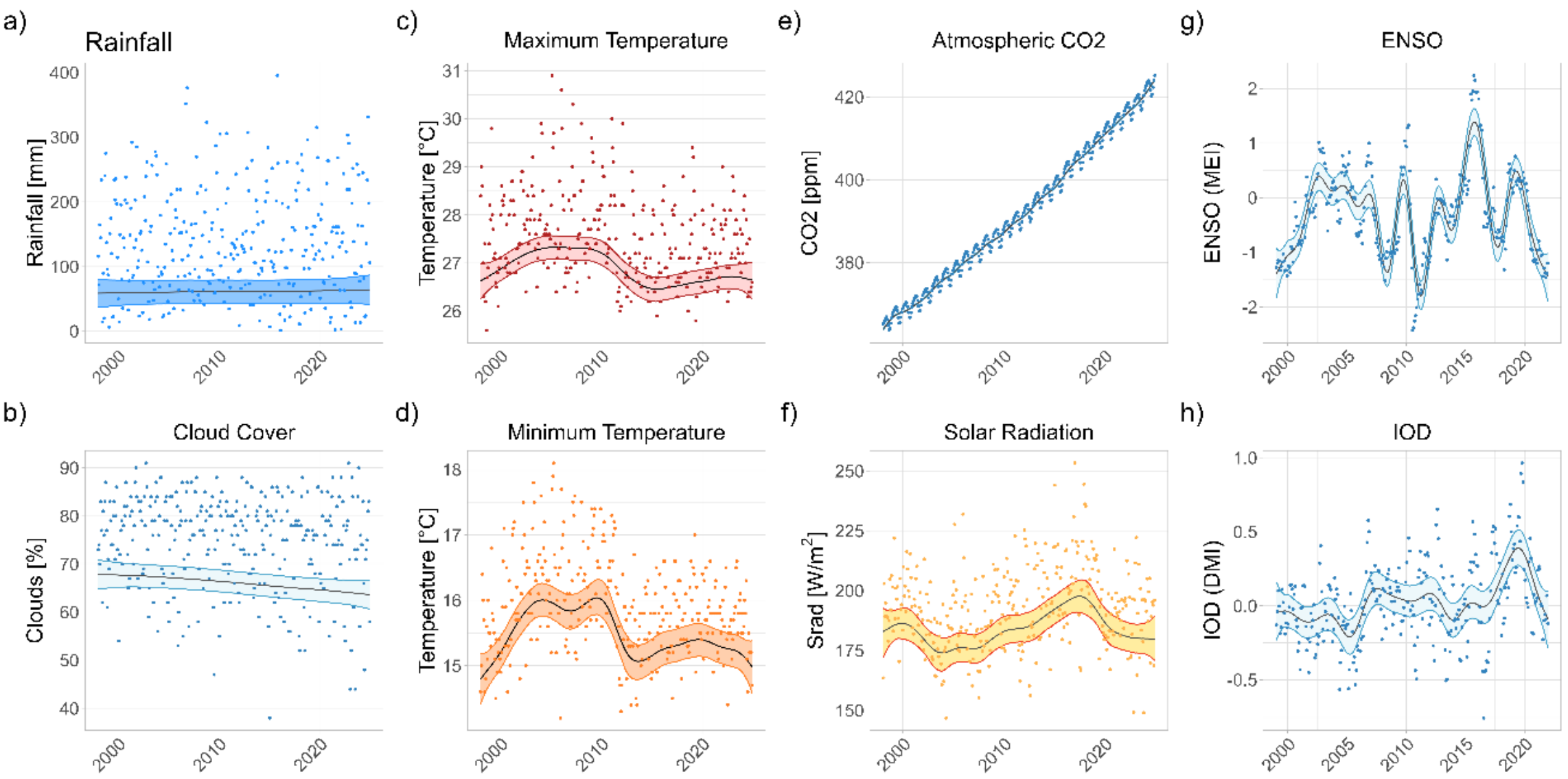

***Figure S2*** *Trend splines of the climate models for (a) rainfall, (b) fractional cloud cover, (c) $T_{max}$, (d) $T_{min}$, (e) atmospheric $CO_2$, (f) solar radiation, (g) ENSO, and (h) IOD. Shaded areas represent the 95% credible interval of the estimated mean. Jittered points show the observed data.*

***Table S1 Posterior estimates of seasonal climate effects on young leaf production.*** *The table presents posterior mean estimates and corresponding 95% credible intervals (l95, u95) for seasonal climate predictors retained in the top-supported model for each species and for the community. Only predictors included in the top model (based on model comparison criteria described in the Methods) are shown. Positive estimates indicate that climate variables are positively associated with increased young leaf production, whereas negative estimates indicate the opposite effect. Credible intervals that do not overlap zero indicate strong statistical support for the effect.*

| *Species* | $CV_{seasonal}$ | *Estimate* | *l95* | *u95* |
|---|---|---|---|---|
| *Community* | Cloud Cover | 0.092 | 0.08 | 0.103 |
| *Community* | Minimum Temperature | 0.014 | 0.003 | 0.024 |
| *Community* | Solar Radiation | 0.039 | 0.028 | 0.049 |
| *Albizia grandibracteata* | Cloud Cover | 0.093 | 0.051 | 0.134 |
| *Albizia grandibracteata* | Minimum Temperature | 0.095 | 0.056 | 0.134 |
| *Albizia grandibracteata* | Solar Radiation | 0.043 | 0.005 | 0.081 |
| *Celtis africana* | Cloud Cover | 0.083 | 0.051 | 0.115 |
| *Celtis africana* | Solar Radiation | 0.006 | -0.026 | 0.039 |
| *Celtis gomphophylla* | Cloud Cover | 0.235 | 0.195 | 0.275 |
| *Celtis gomphophylla* | Minimum Temperature | 0.199 | 0.164 | 0.235 |
| *Celtis gomphophylla* | Solar Radiation | 0.138 | 0.103 | 0.173 |
| *Dombeya kirkii* | Cloud Cover | 0.207 | 0.171 | 0.241 |
| *Dombeya kirkii* | Minimum Temperature | 0.028 | -0.005 | 0.06 |
| *Dombeya kirkii* | Solar Radiation | 0.014 | -0.018 | 0.045 |
| *Funtumia africana* | Cloud Cover | 0.042 | 0.007 | 0.077 |
| *Funtumia africana* | Solar Radiation | 0.076 | 0.041 | 0.112 |
| *Macaranga schweinfurthii* | Rainfall | 0.114 | 0.07 | 0.16 |
| *Macaranga schweinfurthii* | Solar Radiation | -0.048 | -0.093 | -0.003 |
| *Millettia dura* | Cloud Cover | 0.171 | 0.129 | 0.213 |
| *Millettia dura* | Minimum Temperature | 0.045 | 0.006 | 0.083 |
| *Millettia dura* | Solar Radiation | 0.041 | 0.002 | 0.078 |
| *Parinari excelsa* | Minimum Temperature | -0.027 | -0.061 | 0.008 |
| *Parinari excelsa* | Rainfall | -0.019 | -0.054 | 0.016 |
| *Prunus africana* | Minimum Temperature | -0.103 | -0.136 | -0.07 |
| *Prunus africana* | Rainfall | 0.024 | -0.01 | 0.057 |
| *Prunus africana* | Solar Radiation | 0.062 | 0.03 | 0.093 |
| *Trilepisium madagascariense* | Rainfall | 0.045 | 0.013 | 0.077 |
| *Trilepisium madagascariense* | Solar Radiation | 0.041 | 0.009 | 0.074 |
| *Vepris nobilis* | Rainfall | 0.125 | 0.091 | 0.161 |
| *Vepris nobilis* | Solar Radiation | -0.012 | -0.048 | 0.024 |

***Table S2 Posterior estimates of long-term climate effects on young leaf production.*** *The table presents posterior mean estimates and corresponding 95% credible intervals (l95, u95) for long-term climate predictors retained in the top-supported model for each species and for the community. Only predictors included in the top model (based on model comparison criteria described in the Methods) are shown. Positive estimates indicate that climate variables are positively associated with increased young leaf production, whereas negative estimates indicate the opposite effect. Credible intervals that do not overlap zero indicate strong statistical support for the effect.*

| *Species* | $CV_{long\text{-}term}$ | *Estimate* | *l95* | *u95* |
|---|---|---|---|---|
| *Community* | CO2 | 0.199 | 0.188 | 0.209 |
| *Community* | Cloud Cover | 0.034 | 0.024 | 0.044 |
| *Community* | Solar Radiation | 0.077 | 0.067 | 0.088 |
| *Albizia grandibracteata* | Maximum Temperature | -0.113 | -0.152 | -0.075 |
| *Albizia grandibracteata* | Rainfall | 0.052 | 0.017 | 0.087 |
| *Celtis africana* | CO2 | 0.136 | 0.103 | 0.168 |
| *Celtis africana* | Rainfall | -0.044 | -0.074 | -0.013 |
| *Celtis africana* | Solar Radiation | 0.023 | -0.008 | 0.055 |
| *Celtis gomphophylla* | CO2 | 0.098 | 0.063 | 0.133 |
| *Celtis gomphophylla* | Cloud Cover | 0.032 | -0.001 | 0.067 |
| *Celtis gomphophylla* | Solar Radiation | 0.081 | 0.047 | 0.115 |
| *Dombeya kirkii* | CO2 | 0.314 | 0.279 | 0.35 |
| *Dombeya kirkii* | Cloud Cover | 0.085 | 0.052 | 0.119 |
| *Dombeya kirkii* | Solar Radiation | 0.063 | 0.033 | 0.095 |
| *Funtumia africana* | CO2 | 0.225 | 0.192 | 0.259 |
| *Funtumia africana* | Cloud Cover | 0.045 | 0.009 | 0.079 |
| *Funtumia africana* | Solar Radiation | 0.17 | 0.136 | 0.205 |
| *Macaranga schweinfurthii* | Rainfall | 0.049 | 0.005 | 0.094 |
| *Macaranga schweinfurthii* | Solar Radiation | -0.081 | -0.127 | -0.034 |
| *Millettia dura* | Rainfall | 0.04 | 0.004 | 0.077 |
| *Millettia dura* | Solar Radiation | 0.111 | 0.076 | 0.146 |
| *Parinari excelsa* | CO2 | 0.382 | 0.349 | 0.416 |
| *Parinari excelsa* | Rainfall | -0.014 | -0.048 | 0.02 |
| *Parinari excelsa* | Solar Radiation | 0.09 | 0.055 | 0.124 |
| *Prunus africana* | CO2 | 0.14 | 0.109 | 0.172 |
| *Prunus africana* | Cloud Cover | 0.058 | 0.024 | 0.093 |
| *Prunus africana* | Solar Radiation | 0.075 | 0.04 | 0.109 |
| *Strombosia scheffleri* | CO2 | 0.334 | 0.301 | 0.368 |
| *Strombosia scheffleri* | Cloud Cover | 0.031 | -0.004 | 0.068 |
| *Strombosia scheffleri* | Solar Radiation | 0.12 | 0.085 | 0.155 |
| *Trilepisium madagascariense* | CO2 | 0.048 | 0.01 | 0.086 |
| *Trilepisium madagascariense* | Rainfall | 0.018 | -0.015 | 0.05 |
| *Trilepisium madagascariense* | Solar Radiation | 0.051 | 0.018 | 0.086 |
| *Vepris nobilis* | CO2 | 0.229 | 0.191 | 0.268 |
| *Vepris nobilis* | Cloud Cover | 0.068 | 0.028 | 0.109 |
| *Vepris nobilis* | Solar Radiation | 0.099 | 0.059 | 0.137 |

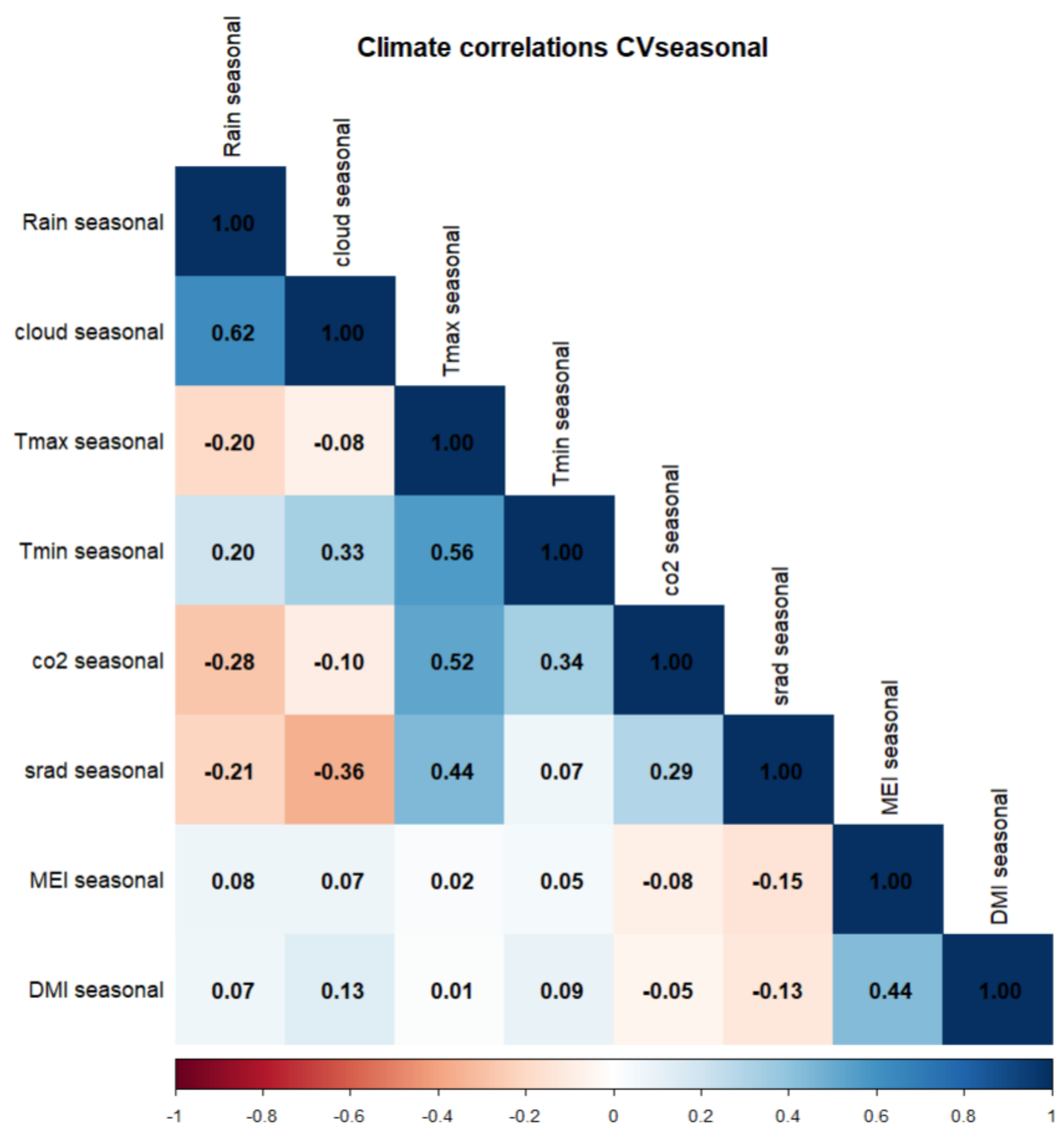

***Figure S3 Pairwise correlations among seasonal climate variability metrics $CVs_{seasonal}$.*** *Pairwise correlation matrix of $CVs_{seasonal}$ for seasonal rainfall, cloud cover, Tmax, Tmin, $CO_2$, solar radiation, MEI, and DMI. Panels show the corresponding Pearson correlation coefficients. The matrix allows visual assessment of collinearity among seasonal climate drivers prior to inclusion in statistical models.*

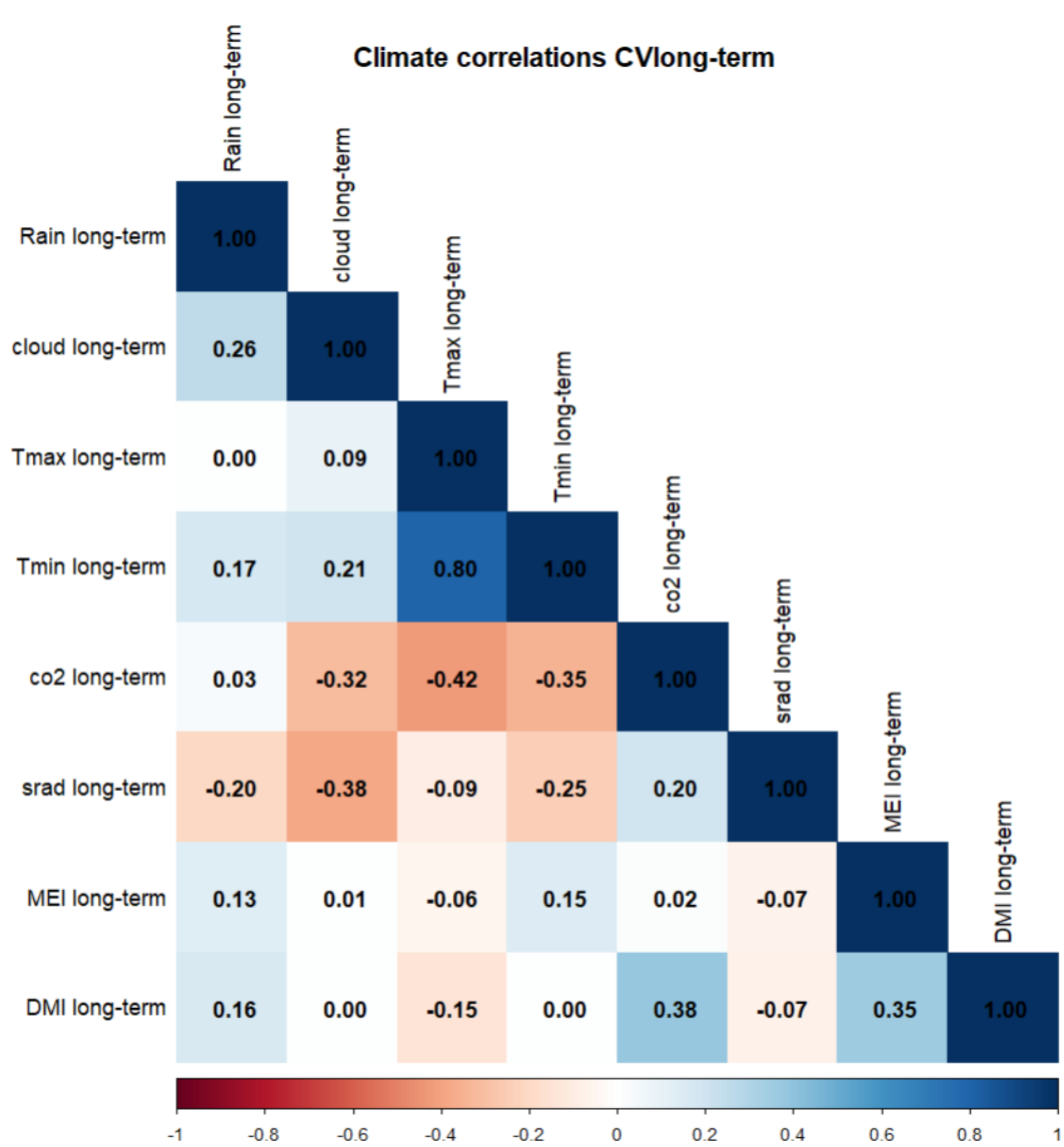

***Figure S4 Pairwise correlations among long-term climate variability metrics $CVs_{long-term}$.*** *Pairwise correlation matrix of $CVs_{long-term}$ for long-term rainfall, cloud cover, Tmax, Tmin, $CO_2$, solar radiation, MEI, and DMI. Panels show the corresponding Pearson correlation coefficients. The matrix allows visual assessment of collinearity among long-term climate drivers prior to inclusion in statistical models.*